\documentclass[astrosymb,twocolumn,preprint]{aastex631}

\usepackage{gensymb}
\usepackage{lmodern}
\usepackage{amsmath}
\usepackage{natbib}
\usepackage[shortlabels]{enumitem}
\usepackage{amssymb}
\usepackage{xfrac}
\usepackage{calc}
\usepackage{booktabs}
\usepackage{hyperref}
\usepackage[nameinlink]{cleveref}
\Crefname{figure}{Figure}{Figures}
\Crefname{section}{Section}{Sections}
\Crefname{subsection}{Section}{Sections}
\Crefname{subsubsection}{Section}{Sections}

\usepackage[T1]{fontenc} % for Icelandic Characters
\usepackage[utf8]{inputenc} % for Icelandic Characters 

\setcitestyle{citesep={;}}
\setcitestyle{aysep={}}
\creflabelformat{equation}{#2\textup{#1}#3}

\newcommand{\RNum}[1]{\uppercase\expandafter{{\scshape\romannumeral #1\relax}}}

\shorttitle{The Young Hot Jupiter Candidate HS Psc b}
\shortauthors{Tran et al.}
\graphicspath{{./}{Figures/}}

\begin{document}

\title{The Epoch of Giant Planet Migration Planet Search Program. II. \\ A Young Hot Jupiter Candidate around the AB Dor Member HS Psc\footnote{Based on observations obtained with the Hobby-Eberly Telescope (HET), which is a joint project of the University of Texas at Austin, the Pennsylvania State University, Ludwig-Maximillians-Universitaet Muenchen, and Georg-August Universitaet Goettingen. The HET is named in honor of its principal benefactors, William P. Hobby and Robert E. Eberly.}}

\correspondingauthor{Quang H. Tran}
\author[0000-0001-6532-6755]{Quang H. Tran}
\email{quangtran@utexas.edu}
\affiliation{Department of Astronomy, The University of Texas at Austin, 2515 Speedway, Stop C1400, Austin, TX 78712, USA}

\author[0000-0003-2649-2288]{Brendan P. Bowler}
\affiliation{Department of Astronomy, The University of Texas at Austin, 2515 Speedway, Stop C1400, Austin, TX 78712, USA}

\author[0000-0001-9662-3496]{William D. Cochran}
\affiliation{McDonald Observatory and Center for Planetary Systems Habitability}
\affiliation{Department of Astronomy, The University of Texas at Austin, 2515 Speedway, Stop C1400, Austin, TX 78712, USA}

\author[0000-0003-1312-9391]{Samuel Halverson}
\affiliation{Jet Propulsion Laboratory, California Institute of Technology, 4800 Oak Grove Drive, Pasadena, CA 91109, USA}

\author[0000-0001-9596-7983]{Suvrath Mahadevan}
\affiliation{Department of Astronomy \& Astrophysics, The Pennsylvania State University, 525 Davey Lab, University Park, PA 16802, USA}
\affiliation{Center for Exoplanets \& Habitable Worlds, University Park, PA 16802, USA}
\affiliation{Penn State Astrobiology Research Center, University Park, PA 16802, USA}

\author[0000-0001-8720-5612]{Joe P.\ Ninan}
\affiliation{Department of Astronomy and Astrophysics, Tata Institute of Fundamental Research, Homi Bhabha Road, Colaba, Mumbai 400005, India}

\author[0000-0003-0149-9678]{Paul Robertson}
\affiliation{Department of Physics \& Astronomy, The University of California, Irvine, Irvine, CA 92697, USA}

\author[0000-0001-7409-5688]{Guðmundur Stefánsson}
\affiliation{NASA Sagan Fellow}
\affiliation{Department of Astrophysical Sciences, Princeton University, 4 Ivy Lane, Princeton, NJ 08540, USA}

\author[0000-0002-4788-8858]{Ryan C. Terrien}
\affiliation{Carleton College, One North College St., Northfield, MN 55057, USA}

\accepted{February 27, 2024}

\begin{abstract}
    We report the discovery of a hot Jupiter candidate orbiting HS Psc, a K7 ($\approx$0.7~$M_\odot$) member of the $\approx$130~Myr AB Doradus moving group. Using radial velocities over 4 years from the Habitable-zone Planet Finder spectrograph at the Hobby-Eberly Telescope, we find a periodic signal at $P_b = 3.986_{-0.003}^{+0.044}$ d. A joint Keplerian and Gaussian process stellar activity model fit to the RVs yields a minimum mass of $m_p \sin i = 1.5_{-0.4}^{+0.6}$ $M_\mathrm{Jup}$. The stellar rotation period is well constrained by the Transiting Exoplanet Survey Satellite light curve ($P_\mathrm{rot} = 1.086 \pm 0.003$ d) and is not an integer harmonic nor alias of the orbital period, supporting the planetary nature of the observed periodicity. HS Psc b joins a small population of young, close-in giant giant planet candidates with robust age and mass constraints and demonstrates that giant planets can either migrate to their close-in orbital separations by 130~Myr or form \textit{in situ}. Given its membership in a young moving group, HS Psc represents an excellent target for follow-up observations to further characterize this young hot Jupiter, refine its orbital properties, and search for additional planets in the system.
\end{abstract}

\section{Introduction\label{sec:intro}} 

The processes by which the closest-in ($ P_\mathrm{orb} \lesssim 10$~d) giant planets arrive at their present-day locations remain challenging to observationally constrain. Several formation and migration mechanisms have been introduced to account for this population of hot Jupiters \citep[HJs;][]{Dawson2018}. These include \textit{in-situ} formation \citep[e.g.,][]{Boley2016, Batygin2016}, disk migration \citep[e.g.,][]{Goldreich1980, Ward1997, Kley2012}, or more ``violent'' scenarios of three-body dynamical interactions such as planet-planet scattering or von Zeipel-Lidov-Kozai interactions coupled with high-eccentricity tidal migration \citep[e.g.,][]{Wu2003, Fabrycky2007, Triaud2010, Naoz2011}. However, as the timescales of these processes ($\approx$1~Myr--1~Gyr) are shorter than the typical ages of observed systems, our knowledge of HJ evolutionary pathways is limited.

Young giant planets are valuable because they provide a means to distinguish among these mechanisms, which operate on different characteristic timescales. However, the detection and characterization of young HJs has been hindered by the presence of stellar activity, making these systems challenging to find and validate. As a result, many proposed young hot Jupiters have been disputed, rejected, or lack independent confirmation \citep[e.g.,][]{Hernan-Obispo2010, Figueira2010, vanEyken2012, Hernan-Obispo2015, Tuomi2018, Carleo2018, Carleo2020, Bouma2020a, Carmona2023}. Notable examples of young HJs include V830 Tau b, TAP 26 b, and CI Tau b \citep{Donati2016, Johns-Krull2016, Yu2017}.

More recently, a growing number of large transiting planets have been found orbiting young stars, such as K2-33 b \citep{David2016, Mann2016}, HIP 67522 b \citep{Rizzuto2020}, and TOI-837 b \citep{Bouma2020b}, with radii consistent with Jupiter-sized planets ($R_p \geq 0.5$~$R_\mathrm{Jup}$). However, these systems lack precise mass measurements. As such, it is difficult to determine whether these systems are truly young HJs or instead are inflated Neptunes that have yet to shrink, either from gravitational contraction as they cool or through atmospheric mass loss \citep[e.g.,][]{Fortney2007, Lopez2012, Owen2013, Gupta2019, Gupta2020}, as has been suggested for systems discovered in young clusters \citep[e.g.,][]{Mann2017, Rizzuto2018}.

Given these challenges and the intrinsically low occurrence rate of HJs at field ages \citep[$1.2 \pm 0.38$\%;][]{Wright2012}, each young system provides valuable information about the origin and dynamical evolution of this class of giant planets. Building a statistically robust sample of young HJs provides clues about the migration history and physical mechanism producing close-in giant planets.

The Epoch of Giant Planet Migration planet search program is an ongoing long-baseline, near-infrared (NIR) precision radial velocity (RV) survey targeting young, nearby Sun-like stars \citep{Tran2021}. Our targets comprise bona fide and high-probability candidate members of ten young moving groups (YMGs)---AB Dor, $\beta$ Pic, Carina, Carina-Near, Octans, Tuc-Hor, Pleiades, 32 Ori, Argus, and Pisces---with ages between 20 Myr and 200 Myr. Known binaries and fast rotators ($v \sin i_* > 35$~km s$^{-1}$) are removed to produce a sample consistent with giant planet RV search programs of older field stars \citep[e.g.,][]{Johnson2010}. By observing in the NIR, as opposed to the optical, our program leverages the wavelength dependence of spot-driven activity signals to minimize RV contributions from astrophysical jitter. This opens the possibility of detecting massive, close-in young planets with RV semi-amplitudes comparable to the reduced activity signals. The goal of this survey is to measure the occurrence rate of close-in giant planets at young ages, compare this frequency with similar measurements at older ages, and constrain the dominant timescale and mechanism of giant planet migration.

Here, we present the discovery of a HJ candidate orbiting the young star HS Psc with the Habitable-zone Planet Finder spectrograph (HPF) at McDonald Observatory's Hobby-Eberly Telescope (HET). In \autoref{sec:hs_psc}, we summarize the physical parameters and previous observations of HS Psc. We describe the Transiting Exoplanet Survey Satellite (TESS) photometry and the HPF RV observations of HS Psc in \autoref{sec:observation}. Characterization of the system, including the host star, and modeling of the HPF RVs is presented in \autoref{sec:analysis}. In \autoref{sec:discussion}, we discuss the implications of this discovery and follow-up observations that would be helpful to further validate and characterize HS Psc b. Finally, we summarize our results in \autoref{sec:summary}.

\begin{deluxetable}{lcc}
\tablecaption{\label{tab:prop} Properties of HS Psc}
\tablehead{\colhead{Property} & \colhead{Value} & \colhead{Ref.}}
\startdata
Gaia DR3 ID & 295713106830535424 & 1 \\
TIC ID & 353804063 & 2 \\
2MASS ID & J01372322+2657119 & 3 \\
$\alpha_{2000.0}$ & 01:37:23.38 & 1 \\
$\delta_{2000.0}$ & +26:57:10.03 & 1 \\
$\mu_{\alpha}$\tablenotemark{a} (mas yr$^{-1}$) & $118.237 \pm 0.022$ & 1 \\
$\mu_{\delta}$ (mas yr$^{-1}$) & $-127.589 \pm 0.020$ & 1 \\
$\pi$ (mas) & $26.486 \pm 0.018$ & 1 \\
Distance (pc) & $37.667_{-0.044}^{+0.038}$ & 1 \\
$\mathrm{RUWE_{DR3}}$ & 1.055 & 1 \\
\hline
$G$ (mag) & $10.292 \pm 0.004$ & 1 \\
$G_\mathrm{RP}$ (mag) & $9.455 \pm 0.007$ & 1 \\
$G_\mathrm{BP}$ (mag) & $11.024 \pm 0.009$ & 1 \\
$B$ (mag) & $12.027 \pm 0.053$ & 4 \\
$V$ (mag) & $10.855 \pm 0.063$ & 4 \\
$J$ (mag) & $8.429 \pm 0.023$ & 3 \\
$H$ (mag) & $7.784 \pm 0.021$ & 3 \\
$K_s$ (mag) & $7.642 \pm 0.027$ & 3 \\
$W_1$ (mag) & $7.515 \pm 0.031$ & 5 \\
$W_2$ (mag) & $7.557 \pm 0.021$ & 5 \\
\hline
SpT & K7V & 6 \\
Mass ($M_\odot$) & $0.69 \pm 0.07$ & 7 \\
Radius ($R_\odot$) & $0.65 \pm 0.07$ & 7 \\
Luminosity $(L_\odot)$ & $0.122 \pm 0.007$ & 7 \\
$T_{\mathrm{eff}}$ (K) & $4203 \pm 116$ & 7 \\
$\log g$ (dex)\tablenotemark{b} & $4.66 \pm 0.03$ & 7 \\
$[\mathrm{Fe}/\mathrm{H}]$ (dex) & $-0.05 \pm 0.09$ & 7 \\
$v \sin i_*$ (km s$^{-1}$) & $29.7 \pm 3.1$ & 7 \\
$v_\mathrm{eq}$ (km s$^{-1}$) & $29.7 \pm 2.8$ & 7 \\
$i_*$ ($\degr$) & $90_{-22}^{+0} \degr$ & 7 \\
$P_\mathrm{rot}$ (d) & $1.086 \pm 0.003$ & 7 \\
Age (Myr) & $133_{-20}^{+15}$ & 8 \\
\enddata
\tablenotetext{a}{Proper motion in R.A. includes a factor of $\cos \delta$.}
\tablenotetext{b}{In cgs units.}
\tablerefs{(1) \citet{GaiaDR3_2022};
(2) \citet{Stassun2019};
(3) \citet{Cutri2003};
(4) \citet{Zacharias2012};
(5) \citet{Cutri2014};
(6) \citet{Bowler2019};
(7) This work;
(8) \citet{Gagne2018b}.}
\end{deluxetable}

\section{HS Psc: A Young Sun-like Star\label{sec:hs_psc}}

HS Psc is a bright \citep[$V = 10.86$~mag;][]{Zacharias2013} K7V star \citep{Bowler2019} in the AB Doradus (AB Dor) Moving Group \citep{Schlieder2010, Malo2013, Gagne2018a}. It has an age of $133_{-20}^{+15}$~Myr from its cluster membership \citep{Gagne2018b}, a parallactic distance of $37.667_{-0.044}^{+0.038}$~pc \citep{GaiaDR3_2022}, and a stellar isochrone-inferred mass of $0.69 \pm 0.07$~$M_\odot$ (\autoref{sec:mass+rad}). \autoref{tab:prop} summarizes the kinematic, photometric, and physical properties of HS Psc.

As a young, bright, nearby star, HS Psc has been observed several times by direct-imaging exoplanet search programs. \citet{Brandt2014} targeted HS Psc with with HiCIAO differential imaging instrument \citep{Suzuki2010} at the Subaru Telescope as part of the Strategic Exploration of Exoplanets and Disks with Subaru (SEEDS) program. They detected a faint point source at a projected distance of approximately 200~AU ($\approx$$5\farcs3$) but did not recover it in follow-up observations. \citet{Bowler2015} also acquired deep observations of HS Psc with Keck/NIRC2 in $K_s$ band, reaching contrasts of $\Delta K_s$  = 9.0~mag at 0$\farcs$5 and $\Delta K_s$ = 11.6~mag at 1$''$. \citet{Bowler2019} obtained shallow optical high-resolution imaging of HS Psc with Robo-AO at the Palomar 1.5~m telescope. No nearby substellar or stellar companions were present within the detection limits of both datasets.

\textit{Gaia} DR3 reports a re-normalised unit weight error (RUWE) of 1.055 for HS Psc \citep{GaiaEDR3_2021, GaiaDR3_2022}. RUWE values greater than $1.4$ indicate that the single-star model is a poor fit to the astrometric solution \citep{Lindegren2018, Stassun2021}, so the RUWE value for HS Psc does not indicate the presence of a binary companion. Furthermore, the RMS of our precise radial velocities over 4 years in \autoref{sec:hpf} implies that we can rule out a close-in massive companion for most orbital orientations. Altogether HS Psc appears to be a single young star with no nearby binary companions in the stellar or brown dwarf (BD) regimes, within the detection limits of current imaging and RV surveys.

\section{Observations\label{sec:observation}}
\subsection{TESS Photometry\label{sec:tess}}

\begin{figure*}[!ht]
	    \centering
	    \includegraphics[width=1.0\linewidth]{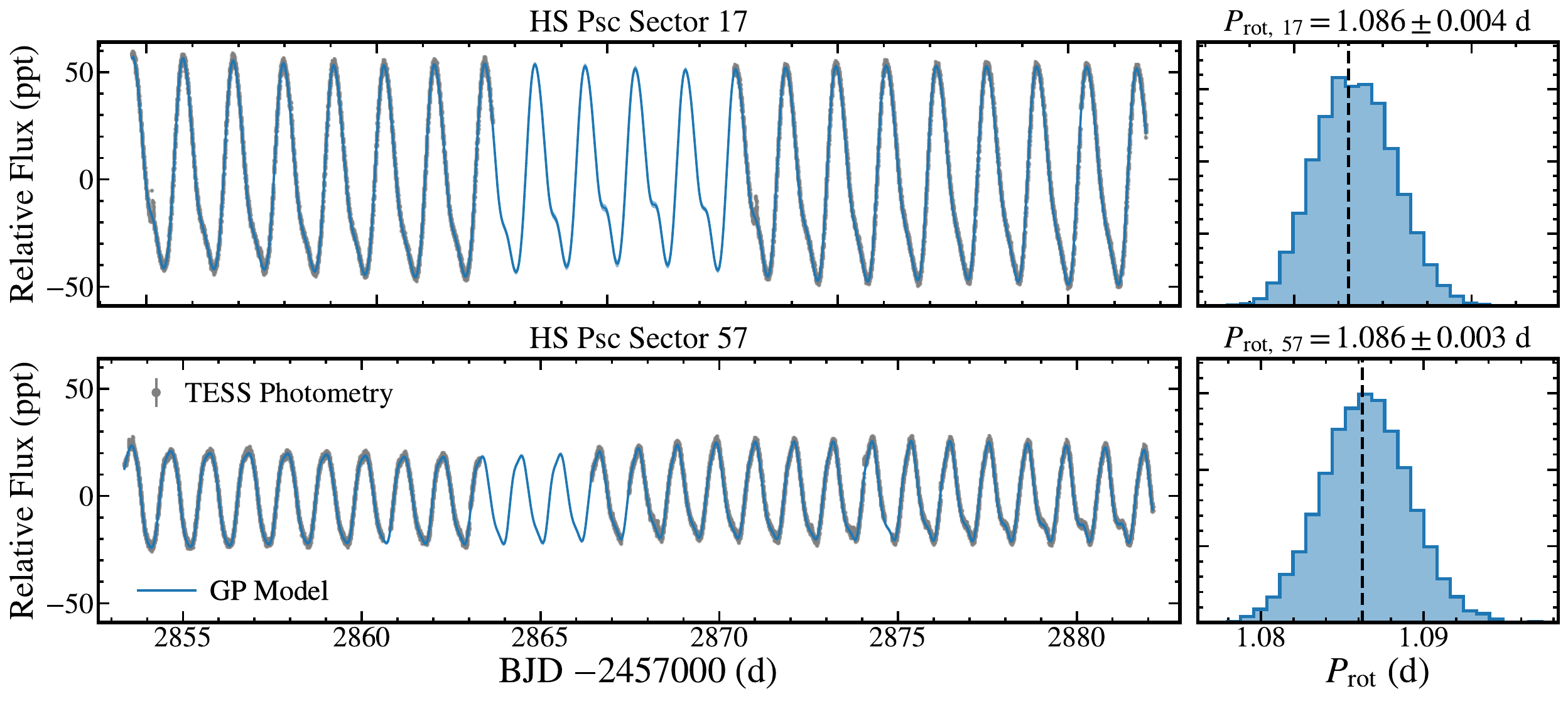}
	    \caption{TESS Sector 17 (top) and 57 (bottom) light curves for HS Psc. The best-fit mean GP model and 1$\sigma$ variance are displayed as solid blue lines and shaded regions, respectively. The posterior distributions and MAP values (dashed vertical line) for the QP GP $P_\mathrm{rot}$ hyperparameter for TESS Sectors 17 and 57 are shown in the right panels. The GP model separately recovers rotation periods of $P_\mathrm{rot, \; 17} = 1.086 \pm 0.003$~d and $P_\mathrm{rot, \; 57} = 1.086 \pm 0.004$~d for Sectors 17 and 57, respectively.}
	    \label{fig:lc+gp}
\end{figure*}

The Transiting Exoplanet Survey Satellite \citep[TESS;][]{Ricker2015} observed HS Psc at 2-minute cadence in Sectors 17 (UT 2019 October 8 to 2019 November 2) and 57 (UT 2022 September 30 to 2022 October 29). The light curve was retrieved and processed following the procedure described in \citet{Bowler2023}. We downloaded the Science Processing Operations Center \citep[SPOC;][]{Jenkins2016} Pre-search Data Conditioning Simple Aperture Photometry (PDCSAP) light curve \citep{Smith2012, Stumpe2012, Stumpe2014} from the MAST data archive\footnote{\href{https://archive.stsci.edu/missions-and-data/tess}{https://archive.stsci.edu/missions-and-data/tess.} All TESS data used is available at MAST:\dataset[10.17909/t9-nmc8-f686]{http://dx.doi.org/10.17909/t9-nmc8-f686} \citep{doi.org10.17909/t9-nmc8-f686}.} using the \texttt{lightkurve} \citep{Lightkurve2018} software package. All photometric measurements flagged as poor (\texttt{DQUALITY} $> 0$) or are listed as \texttt{NaN} are removed. Each TESS sector is first median-normalized and then stitched together. Outlier points from flares and artifacts are removed by flattening the light curve using a high-pass Savitzky-Golay filter \citep{Savitzky1964} and excluding all data outside of three standard deviations of the flattened light curve. The TESS light curve for Sectors 17 and 57 are displayed in the left panels of \autoref{fig:lc+gp}.

\subsection{HPF Spectroscopy\label{sec:hpf}}

As part of our initial results from the Epoch of Giant Planet Migration program, \citet{Tran2021} measured an RV rms of over 200 m s$^{-1}$ for HS Psc with 8 RV epochs. This system was an outlier among our sample of young stars, which had a median RV rms of only 34.3 m s$^{-1}$. As this anomalously high RV scatter could potentially be explained by a close-in giant planet, we increased our monitoring cadence of this system over the following 3 years.

We used HPF on the HET \citep{Ramsey1998, Hill2021} to obtain a total of 83 spectra of HS Psc between UT 2018 November and UT 2022 November. HPF is an NIR (0.81--1.27~$\mu$m), fiber-fed \citep{Kanodia2018b}, environment-stabilized \citep{Mahadevan2012, Mahadevan2014, Stefansson2016} high-resolution \'echelle spectrograph. Wavelength calibration is achieved with a laser frequency comb \citep{Metcalf2019}. HPF has 28 spectral orders with an average resolving power of $R \equiv \lambda$/$\Delta \lambda$ = 55,000 \citep{Mahadevan2012, Ninan2019}.

Observations are obtained and RVs are measured following the procedures detailed in \citet{Tran2021}. HET's flexible scheduling system allows observations to be taken in queue mode \citep{Shetrone2007}. During each epoch, 3 contiguous exposures are obtained to sample high frequency variations. All exposures have a 300~s integration time and the average S/N at $1.07~\mu$m is $115 \pm 30$ per pixel. 1D spectra are optimally extracted using the custom HPF data-reduction pipeline following the procedures described in \citet{Ninan2018}, \citet{Kaplan2019}, and \citet{Metcalf2019}.

Relative RVs, differential line-widths (dLWs), and chromatic indices (CRXs) are measured using a custom least-squares matching pipeline based on the SpEctrum Radial Velocity AnaLyser (\texttt{SERVAL}) code \citep{Zechmeister2018, Tran2021}. Furthermore, the line indices of the \ion{Calcium}{2} infrared triplet (\ion{Ca}{2} IRT) emission lines are measured following \citet{Stefansson2020b}. We use the 8 \'echelle orders corresponding to wavelength ranges in the $z$-band (8535--8895~\AA) and $Y$-band (9933--10767~\AA) for the RV extraction to avoid contamination of strong telluric absorption \citep[orders 4, 5, 6, 14, 15, 16, 17, 18;][]{Tran2021}. The RV and activity indicator measurements are reported in \autoref{tab:hpf_measurements} in \autoref{sec:appendix_measurements}.

From UT 2022 May to June, HPF underwent a vacuum warm-up and cool-down cycle so maintenance work could be carried out on the detector. This has introduced a velocity offset in the RV time series. To account for this unknown offset, RVs obtained prior to and after this downtime are treated as if they are measurements from two different instruments when searching for planets and modeling Keplerian signals. In this work, we label the RV measurements prior to this maintenance ``pre-warmup'' HPF$_1$ and measurements after ``post-warmup'' HPF$_2$ to distinguish between the two observing periods. Altogether there are $N_{\mathrm{RV}, \; \mathrm{HPF}_{1}} = 62$ and $N_{\mathrm{RV}, \; \mathrm{HPF}_{2}} = 21$ observations associated with HPF$_1$ and HPF$_2$, respectively. The pre-warmup RVs have an rms of $\mathrm{RV}_{\mathrm{rms}, \; \mathrm{HPF}_{1}} = 262$~m s$^{-1}$ and an average RV measurement uncertainty of $\bar{\sigma}_{\mathrm{RV}, \; \mathrm{HPF}_1}= 124$~m s$^{-1}$. The post-warmup RVs have an rms of $\mathrm{RV}_{\mathrm{rms}, \; \mathrm{HPF}_{2}} = 264$~m s$^{-1}$ and an average RV measurement uncertainty of $\bar{\sigma}_{\mathrm{RV}, \; \mathrm{HPF}_2}= 150$~m s$^{-1}$. These large measurement uncertainties are primarily driven by the large rotational velocity (see \autoref{sec:spec_params}) of HS Psc.

\section{Results\label{sec:analysis}} 
\subsection{Spectroscopic Parameters\label{sec:spec_params}}

We determine the spectroscopic properties of HS Psc following the empirical spectral matching procedure as described in \citet{Stefansson2020a}, which is based on the \texttt{SpecMatch-Emp} algorithm of \citet{Yee2017}. This technique compares a science spectrum with a library of high-S/N spectra of slowly rotating stars and identifies the best match between the stellar and library spectra. During each fit, the reference star spectrum is convolved with a broadening kernel \citep{Gray1992} in order to estimate the rotational velocity. The final values are estimated from a composite spectra of the five best-fitting reference spectra.

Using the highest-S/N HPF (S/N = 150) spectrum as the science target, we measure the effective temperature ($T_\mathrm{eff}$), metallicity ([Fe/H]), surface gravity ($\log g$), and projected rotational velocity ($v \sin i_*$) of HS Psc. The procedure is applied to the same 8 HPF \'echelle orders used for RV extraction. Uncertainties on $T_\mathrm{eff}$, [Fe/H], and $\log g$ are calculated for each \'echelle order using a cross-validation method following \citet{Stefansson2020a}, which iteratively removes each library spectrum from the sample, fits for $T_\mathrm{eff}$, [Fe/H], and $\log g$ using the interpolated grid, and compares the best-fit measurement to the known library value. The standard deviation of the residuals between the recovered and library values is adopted as the uncertainty of the parameters for each \'echelle order. The weighted mean and weighted standard deviation of the spectral order measurements are used as the final values and uncertainties for each parameter. This cross-validation method cannot determine uncertainties on $v \sin i_*$ as all library spectra are of slowly rotating stars. Thus, for $v \sin i_*$, we adopt the median value and standard deviation of the 8 \'echelle order values. We find an effective temperature of $T_\mathrm{eff} = 4203 \pm 116$~K, a metallicity of $\mathrm{[Fe/H]} = -0.05 \pm 0.09$~dex, a surface gravity of $\log g = 4.66 \pm 0.03$~dex, and a stellar rotational velocity of $v \sin i_* = 29.7 \pm 3.1$~km s$^{-1}$ for HS Psc.

These values are consistent with measurements reported in the literature. \citet{McCarthy2012} found $T_\mathrm{eff} = 4400 \pm 105$ K using low-resolution ($R \sim 3575$) optical spectra. \citet{Stassun2019} report an effective temperature of $T_\mathrm{eff} = 4215 \pm 128$~K and a surface gravity of $\log g = 4.64 \pm 0.11$~dex. \citet{Barenfeld2013} found a near-Solar representative metallicity for the AB Dor moving group of $\mathrm{[Fe/H]} = 0.02 \pm 0.02$~dex.

\citet{Schlieder2010} report a much lower rotational velocity of $v \sin i_* = 10 \pm 2$ km s$^{-1}$ using spectra from CSHELL \citep{Greene1993}. This discrepancy between $v \sin i_*$ measured using our HPF and their CSHELL spectra may be due to the lower resolution of CSHELL, which can be as low as $R =$ 5,000, depending on the slit choice. In \citet{Tran2021}, we previously measured a relatively uncertain projected rotational velocity of $v \sin i_* = 22 \pm 8$~km s$^{-1}$. Our new measurement of $v \sin i_* = 29.7 \pm 3.1$~km s$^{-1}$ is more precise and based on empirical templates instead of synthetic models.

\subsection{Stellar Mass and Radius\label{sec:mass+rad}} 

We estimate the stellar radius by fitting the spectral energy distribution (SED) of HS Psc using the open software package \texttt{ARIADNE} \citep{Vines2022}. \texttt{ARIADNE} employs a Bayesian model averaging approach to convolve four stellar atmosphere model grids---PHOENIX v2 \citep{Husser2013}, BT-Settl \citep{Allard2011}, \citet{Kurucz1993}, and \citet{Castelli2003}---with response functions of common broadband photometric filters. The model leaves distance, stellar radius, line-of-sight extinction ($A_V$), and excess photometric uncertainty terms as free parameters. Synthetic SEDs are generated by interpolating the model grids in $T_\mathrm{eff}$--$\log g$--$\mathrm{[Fe/H]}$ space and fit to available photometry. For our fitting process, we use the $B$, $V$, $G_\mathrm{BP}$, $G$, $G_\mathrm{RP}$, $J$, $H$, $K_s$, $W_1$, and $W_2$ bandpasses.

For $T_\mathrm{eff}$, $\log g$, and [Fe/H], we adopt Gaussian priors with means and standard deviations set to the values and uncertainties reported in \autoref{sec:spec_params}. The distance prior uses a Gaussian distribution with the mean and standard deviation set to the Gaia DR3 distance estimate and the larger of the upper and lower uncertainties, $\mathcal{N}(37.667, 0.044)$\footnote{$\mathcal{N}(a, b)$ refers to the normal distribution with a mean of $a$ and standard deviation of $b$.}~pc, the stellar radius prior set to a Gaussian distribution with the mean and standard deviation set to the Gaia DR3 Final Luminosity Age Mass Estimator \citep[FLAME;][]{Gaia_DR2} derived radius and uncertainty, $\mathcal{P}(R_*) = \mathcal{N}(0.6561, 0.019)$~$R_\odot$, and the line-of-sight extinction prior set to a uniform distribution with an upper limit set to the maximum line-of-sight ($A_{V, \; \mathrm{max}} = 0.235$~mag) extinction from the \citet[][SFD]{Schlegel1998} galactic dust map \citep{Schlafly2011} as determined by the software package \texttt{dustmaps} \citep{Green2018}. The excess photometric noise terms all have Gaussian priors centered at 0 with a standard deviation equal to 10 times the photometric uncertainty of each band.

The posterior distributions of all fitted parameters in the model are sampled using the dynamic nested sampler \texttt{dynesty} \citep{Skilling2004, Skilling2006, Speagle2020, Koposov2023}. The sampling is initialized with 5000 live points and terminates when the evidence tolerance reaches a threshold of $\texttt{dlogz} < 0.1$. From this fit to the SED, we infer a stellar radius of $R_* = 0.65 \pm 0.01$~$R_\odot$. The parameters derived from the grid-interpolation are $T_\mathrm{eff} = 4232_{-38}^{+45}$~K, $\log g = 4.66 \pm 0.03$~dex, and $\mathrm{[Fe/H]} = -0.08_{-0.07}^{+0.08}$~dex, which are consistent with the values measured in \autoref{sec:spec_params}. The bolometric luminosity computed from the SED fitting procedure is $L_* = 0.121 \pm 0.006$~$L_\odot$. Using the derived radius and surface gravity, we estimate a mass of $M_* = 0.70 \pm 0.05$~$M_\odot$ for HS Psc. However, this mass is sensitive to the inferred surface gravity.

We infer the mass of HS Psc using the open software package \texttt{isochrones} \citep{Morton2015}. \texttt{isochrones} determines fundamental stellar properties by fitting combinations of photometric bandpasses and physical values to synthetic values generated using interpolated grids of evolutionary models. For the fitting routine, we utilize the MESA Isochrones and Stellar Tracks \citep[MIST;][]{Dotter2016, Choi2016} evolutionary model grids, all broadband photometry listed in \autoref{tab:prop}, all parameters measured in \autoref{sec:spec_params}, and the Gaia DR3 parallax.

As \texttt{isochrones} samples the age parameter as $\log_{10}(\mathrm{age})$, we set a flat prior in log-space based on the AB Dor moving group age, $\mathcal{P}(\tau_*) = \log \mathcal{U}[113, 148]$\footnote{$\mathcal{U}[a, b]$ refers to a uniform distribution bounded by and inclusive of lower limit $a$ and upper limit $b$.}~Myr. We also adopt a Gaussian prior based on the representative metallicity of the AB Dor moving group from \citet{Barenfeld2013}, $\mathcal{P}(\mathrm{[Fe/H]}) = \mathcal{N}(0.02, 0.02)$~dex. For the distance and extinction priors, we adopt the same priors as in the SED fitting, $\mathcal{P}(\mathrm{distance}) = \mathcal{N}(37.667, 0.044)$~pc and $\mathcal{P}(A_V) = \mathcal{U}[0, 0.235]$~mag. Following the default distribution adopted by \texttt{isochrones}, the mass prior follows the log-normal initial mass function from \citet[][Equation 17]{Chabrier2003}.

The posterior distributions of all fit and derived parameters are sampled with the multimodal nested sampling algorithm \texttt{MultiNest} \citep{Feroz2009, Feroz2019} using the open software package \texttt{PyMultiNest} \citep{Buchner2014}. We initialize the sampling with 5000 live points. The best-fit stellar mass and radius are $M_* = 0.686 \pm 0.003$~$M_\odot$ and $R_* = 0.628 \pm 0.002$~$R_\odot$. The best-fit inferred parameters are $T_\mathrm{eff} = 4338 \pm 9$~K, $\log g = 4.679 \pm 0.001$~dex, and $\mathrm{[Fe/H]} = 0.10 \pm 0.02$~dex, which are consistent with the measurements adopted in \autoref{sec:spec_params} to within 2$\sigma$. Altering the age, metallicity, and distance priors do not substantially change these results.

As a comparison to these measurements, we also report masses and radii estimated using relationships with other parameters such as the effective temperature, metallicity, and luminosity. Combining the parameters derived in \autoref{sec:spec_params} with the empirical functions calibrated using 190 stars from \citet{Torres2010} relating $T_\mathrm{eff}$, $\log g$, and [Fe/H] to $M_*$ and $R_*$ yields a stellar mass and radius of $M_* = 0.60 \pm 0.04$~$M_\odot$ and $R_* = 0.59 \pm 0.02$~$R_\odot$. Similarly, using the Stefan–Boltzmann law, the SED-computed luminosity ($L_* = 0.121 \pm 0.006$~$L_\odot$), and our adopted effective temperature ($T_\mathrm{eff} = 4203 \pm 116$ K) returns a radius of $R_* = 0.66 \pm 0.04$~$R_\odot$. \citet{Stassun2019} estimates a mass and radius of $M_* = 0.66 \pm 0.08$~$M_\odot$ and $R_* = 0.647 \pm 0.062$~$R_\odot$, respectively, for HS Psc.

We adopt the stellar radius and mass inferred with \texttt{ARIADNE} and \texttt{isochrones}, respectively. Given the range of inferred masses and radii in our analysis and compared with similar dispersion in estimates for other K7V members of the AB Dor moving group, we conservatively adopt a flat 10\% estimate for the uncertainty in both parameters. This is larger than the characteristic scatter of inferred masses and radii for comparatively old main-sequence stars \citep[$\sim$5\%;][]{Tayar2022}. The final adopted stellar mass and radius for HS Psc is $M_* = 0.69 \pm 0.07 \; M_\odot$ and $R_* = 0.65 \pm 0.07 \; R_\odot$.

\subsection{Rotation Period, Stellar Inclination, and Transit Search\label{sec:rot_per}} 

The TESS light curve of HS Psc exhibits clear and consistent modulation at the levels of 5\% and 2.5\% during Sectors 17 and 57, respectively. While the amplitude (and to a lesser degree the phase) of the signal changes between the two observational windows, its periodicity does not. This suggests that the observed variability is driven by long-lived spots on the stellar surface and therefore can be used to accurately estimate the stellar rotation period.

The rotation period is derived by independently fitting a quasi-periodic (QP) Gaussian process (GP) to each TESS Sector light curve. QP GPs have been shown to accurately infer stellar rotation periods from photometric time series \citep[e.g.,][]{Angus2018}. The TESS Sectors are treated separately as significant evolution has occurred in the 2.91 years between Sectors 17 and 57 (\autoref{fig:lc+gp}). We bin the TESS photometry to 30-minute cadence to improve computational efficiency while retaining large-scale, activity-driven structure. We use a QP kernel of the form:

\begin{equation} \label{eq:quasi_per_kernel_lc}
    k_\mathrm{QP}(\tau) = A^2 \: \mathrm{exp}\left( -\frac{\tau^2}{2 l^2} - \frac{2 \: \mathrm{sin}^2\left(\frac{\pi \tau}{P_\mathrm{rot}}\right)}{\theta^2} \right),
\end{equation}

\noindent{}where $\tau$ is the time interval between any two points in time $t_i$ and $t_j$, $\lvert t_i - t_j \rvert$, $A$ is the amplitude, $l$ is the local correlation timescale, $P_\mathrm{rot}$ is stellar rotation period, and $\theta$ is the smoothness of the periodic correlation.

We use the \texttt{emcee} open Python package \citep{Foreman-Mackey2013, Foreman-Mackey2019} to sample the posteriors of the kernel hyperparameters and an additional white-noise jitter term, $\sigma_\mathrm{jit}$. We impose non-informative uniform priors for each hyperparameter; for $A$, $l$, $\theta$, $\sigma_\mathrm{jit}$, and $P_\mathrm{rot}$, we adopt $\mathcal{P}(A) = \mathcal{U}[0.01, 5000]$~ppt, $\mathcal{P}(l) = \mathcal{U}[0.001, 10]$, $\mathcal{P}(\theta) = \mathcal{U}[0.001, 10]$, $\mathcal{P}(\sigma_\mathrm{jit}) = \mathcal{U}[\log(0.001), \log(100)]$~ppt, and $\mathcal{P}(P_\mathrm{rot}) = \mathcal{U}[0.1, 10]$~d, respectively. Sampling is initialized with 50 walkers and 10,000 steps, for a total of $5 \times 10^6$ samples, the first 50\% are removed as burn-in, and convergence is confirmed using the autocorrelation time. We adopt the $P_\mathrm{rot}$ MAP value and standard deviation of the posteriors as the rotation period measurement and uncertainty. We find separate rotation periods of $P_\mathrm{rot, \; 17} = 1.086 \pm 0.003$~d and $P_\mathrm{rot, \; 57} = 1.086 \pm 0.004$~d for TESS Sectors 17 and 57, respectively.

These rotation periods are consistent with the values derived from the SuperWASP \citep[$P_\mathrm{rot} = 1.0852$ d;][]{Norton2007} and KELT \citep[$P_\mathrm{rot} = 1.0859$ d;][]{Oelkers2018} photometric surveys. Thus, we safely adopt the weighted mean and mean error of the two TESS Sectors, $P_\mathrm{rot} = 1.086 \pm 0.003$ d, as the stellar rotation period of HS Psc. The processed TESS light curves and marginalized $P_\mathrm{rot}$ posteriors for Sectors 17 and 57 are displayed in \autoref{fig:lc+gp}.

Using the rotation period, stellar radius, and projected rotational velocity, we infer the line-of-sight stellar inclination, $i_*$, following the Bayesian framework from \citet{Masuda2020}. We utilize the analytical expressions described in \citet[][Equation 9]{Bowler2023} to obtain the posterior distribution of $i_*$ while accounting for the correlation between the projected and equatorial rotational velocities. Here, we adopt the TESS rotation period of $P_\mathrm{rot} = 1.086 \pm 0.003$~d, a stellar radius of $R_* = 0.65 \pm 0.07$~$R_\odot$, and a projected rotational velocity of $v \sin i_* = 29.7 \pm 3.1$~km s$^{-1}$. It is evident, given the projected rotational velocity, $v \sin i_* = 29.7 \pm 3.1$~km s$^{-1}$, and the equatorial velocity, $2 \pi R_* / P_\mathrm{rot} = v_\mathrm{eq} = 29.7 \pm 2.8$~km s$^{-1}$, that HS Psc is seen approximately equator-on. We find a maximum \textit{a posteriori} (MAP) value of $83\degr$, a 68\% credible interval spanning 67--90$\degr$, and a 95\% credible interval spanning 52--90$\degr$ for the stellar inclination.

As a close-in giant planet, HS Psc b's geometric transit probability is relatively high at $\approx$6\%. We use the Notch filter and Locally Optimized Combination of Rotations (LOCoR) detrending algorithms \citep{Rizzuto2017} to search for potential transit events in the TESS light curve. The Notch and LOCoR filters use a combination of a transit-shaped notch and quadratic continuum to search for transit-like events in small windows of the light curve. These algorithms have been used to discover planets smaller than Jupiter transiting young ($\tau_* < 100$ Myr) stars \citep[e.g.,][]{Rizzuto2018, Newton2019, Rizzuto2020}.

We run the Notch filter to detrend the original TESS light curve\footnote{Transit searches are conducted on PDCSAP light curve prior to normalization and sigma-clipping as described in \autoref{sec:tess}.} and, as HS Psc is a fast rotator, we apply the LOCoR algorithm to account for aliases of its rotation period ($P_\mathrm{rot} = 1.086 \pm 0.003$~d). Using various window sizes (0.05--0.5~d), we do not detect a signal at the periodicity of the planet ($P_b = 3.986$~d) in the detrended light curves. Several strong signals do emerge at other periodicities, but inspection of the phase-folded, detrended light curves at those observed periods and times of inferior conjunction indicate that these signals are artifacts of the detrending process or remnants of stellar activity that were not detrended.

\begin{figure}[!t]
	    \centering
	    \includegraphics[width=1.0\linewidth]{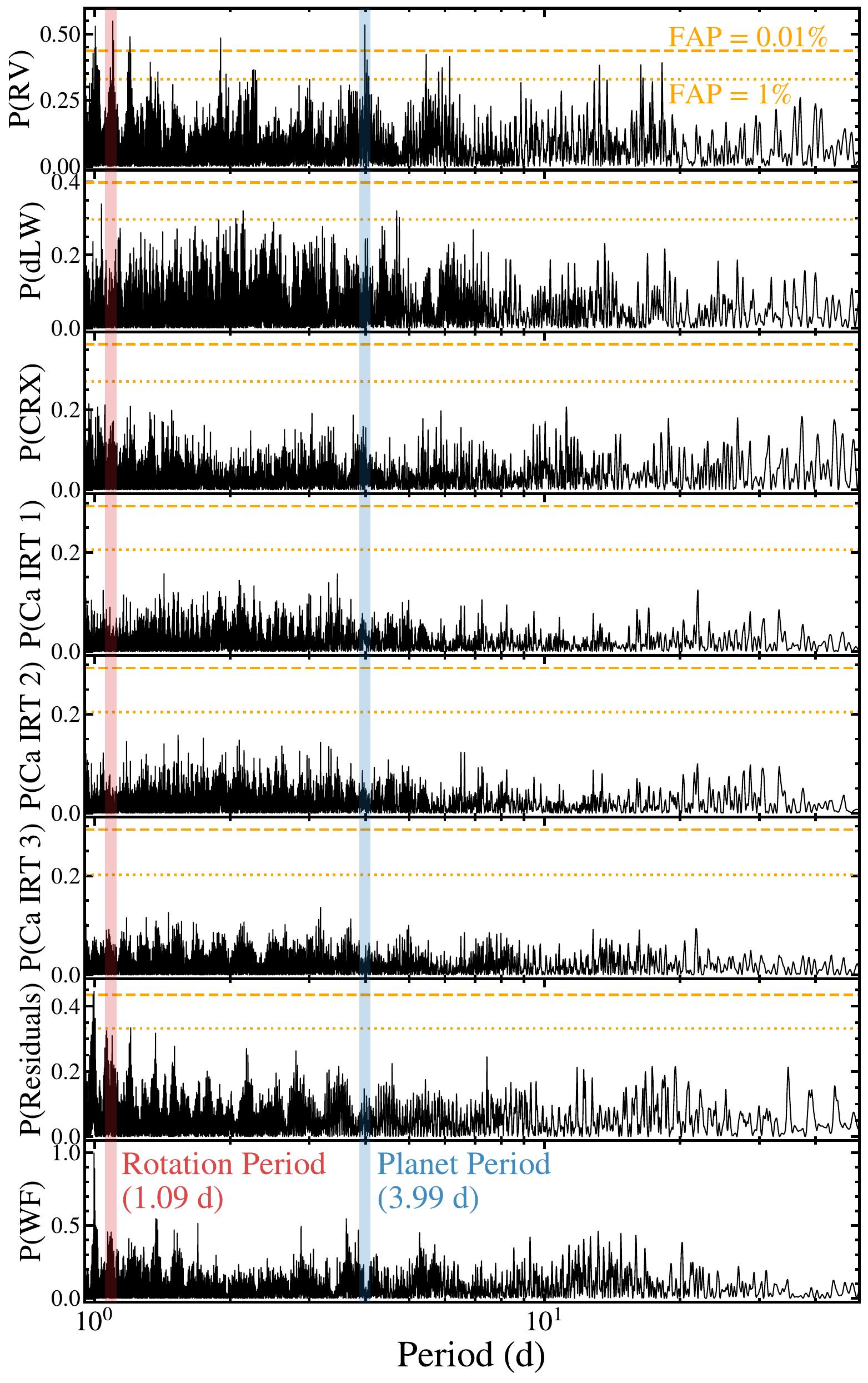}
	    \caption{Generalized Lomb-Scargle periodograms of the spectral time series. Beginning with the top panel, the periodograms are calculated for the HPF RVs, the differential line width (dLW) activity indicator, the chromatic index activity indicator (CRX), the \ion{Ca}{2} IRT indices for the \ion{Ca}{2} 1, 2, and 3 IRT emission lines, the RV residuals after subtracting the best-fit planetary model, and the spectral window function (WF). The planet orbital period ($P_b = 3.99$~d) and the stellar rotation period ($P_\mathrm{rot} = 1.086$~d) are highlighted in the blue and red vertical shaded regions, respectively. False alarm probabilities (FAPs) of 1\% and 0.01\%, calculated using the bootstrap method, are shown as orange dotted and dashed lines, respectively. There are five peaks above the FAP = 0.01\% threshold in the periodogram of the RVs; these represent the nightly sampling alias (seen at 1.0 day in the bottom panel with the WF periodogram), the stellar rotation period, the orbital period, and likely aliases of the stellar rotation and orbital periods.}
	    \label{fig:periodograms}
\end{figure}

\subsection{Periodogram and Stellar Activity Analysis}

To search for periodic signals from planets and investigate the stellar activity of HS Psc, we compute the GLS periodogram of the RV observations and associated activity indicators over the frequency range 0.0005--1.0526~d$^{-1}$ (0.95--2000~d). This upper frequency limit near 1 day corresponds to the first peak of the spectral window function, which is caused by a strong sampling alias resulting from nightly observations frequently seen in ground-based observations. This frequency represents the lowest limit at which the data can reliably identify periodicities. The periodograms are calculated on the concatenated sets of measurements (HPF$_1$ and HPF$_2$).

To test if the choice to combine HPF$_1$ and HPF$_2$ measurements impacts our results, we calculate the GLS periodogram with both the \texttt{LombScargle} and the \texttt{LombScargleMultiband} classes in the open-source Python package \texttt{astropy.timeseries} \citep{Astropy2013, Astropy2018, Astropy2022}, which are designed to treat single time series and multiband time series, respectively. We also calculate GLS periodograms on the concatenated RVs after subtracting the average of each dataset and after subtracting the derived RV offsets. Furthermore, we also calculate periodograms on the RV data separated by even and odd observations and in two halves. We find that all periodograms are similar to each other and periodogram features remain robust against these changes. We thus safely choose to calculate the single-band periodogram on the concatenated measurements for further analysis.

\autoref{fig:periodograms} displays the GLS periodograms for the RV observations, the associated activity indicators (dLWs, CRXs, line indices of the \ion{Ca}{2} IRT), the RV residuals after subtracting the best-fit planetary model (\autoref{sec:rv_fit}), and the spectral window function over the period range of 0.95--50~d. The FAPs at the 1\% and 0.01\% levels are calculated using the \texttt{bootstrap} method and are shown in dotted and dashed orange lines, respectively.

Five peaks rise above the 0.01\% threshold in the RV periodogram. A peak at 1.003~d corresponds to an alias caused by nightly sampling, which is also seen in the window function periodogram. Significant peaks appear near the rotation period (1.09~d) and a likely alias sequence of the rotation period ($\sim$1.2~d). Finally, two significant peaks are observed at 3.99~d and a possible harmonic alias of this signal, 1.91~d. We interpret the $\approx$4-day signal as a planet candidate, as no significant peaks emerge in the activity indicators at or near this frequency. No additional periodic signals that may have been overwhelmed by the initial planetary signal emerge in the periodogram of the residuals; only the $\sim$1-day synoptic nightly sampling alias remains significant at the FAP = 0.01\% level.

To further test for correlations between the RVs and associated spectral indicators, we compute the Pearson's correlation coefficient (Pearson's $r$) and $p$-values between the RVs and activity indicators (see \autoref{sec:hpf}). This test is conducted as spectral indicators that trace distortions in the spectral line profile induced by starspots may correlate with RV measurements \citep[e.g.,][]{Queloz2001, Boisse2011, Meunier2013}. Here, $r$ values measure correlation strength, ranging from $-1$ for perfect anti-correlation and $1$ for perfect correlation. $p$-values report the probability of observing a corresponding $r$ value if the data are not correlated, so a lower $p$-value represents a higher probability that the data are correlated. For the HPF RVs and associated activity indicators, all Pearson's $r$ values have an absolute value less than 0.5 and all $p$-values are greater than $0.05$, indicating that none of the activity indicators are significantly correlated with the HPF RVs. The maximum correlation coefficient is $r = 0.4$, with a $p$-value of 0.07, for the HPF$_2$ RVs and the corresponding \ion{Ca}{2} IRT 2 line index. The other \ion{Ca}{2} IRT indices have similar $r$ values.

\subsection{Radial Velocities and Gaussian Process Modeling\label{sec:rv_fit}}

The youth and persistent photometric variability of HS Psc signifies that strong, correlated starspot-driven modulations should be present in the RV observations of HS Psc b. Red-noise GP models are now commonly employed to fit RV variability arising from stellar activity \citep[e.g.,][]{Affer2016, Damasso2017, Faria2020, Stefansson2022}. To take into account the expected correlation of RV signals caused by stellar activity, we perform three separate model fits to the HPF RVs that vary in their treatment of stellar activity:
\begin{enumerate}[label={\textbf{Model \theenumi.}},itemsep=0ex,partopsep=1ex,parsep=0ex,labelindent=12pt,labelwidth=\widthof{Model\theenumi.}+\labelsep,leftmargin=!]
    \item Keplerian-only,
    \item Keplerian and Quasi-periodic GP,
    \item Keplerian and Mat\'ern--5/2 GP.
\end{enumerate}
Model fits are conducted with all 83 HPF RVs using the \texttt{pyaneti} modeling suite \citep{Barragan2019, Barragan2022}.

All three models fit for a Keplerian orbit with 5 parameters: orbital period ($P_b$), time of inferior conjunction ($T_{0, b}$), RV semi-amplitude ($K_b$), and the parameterized forms of eccentricity and argument of periastron ($\sqrt{e_b} \sin \omega_*$ and $\sqrt{e_b} \cos \omega_*$). Furthermore, for each set of HPF RV observations, we add a ``jitter'' term ($\sigma_{\mathrm{HPF}_{1}}$ and $\sigma_{\mathrm{HPF}_{2}}$) to account for any additional instrumental noise not represented in the measurement uncertainties, as well as zero-point velocity terms ($\gamma_{\mathrm{HPF}_{1}}$ and $\gamma_{\mathrm{HPF}_{2}}$) to account for any systematic offsets. Models 2 and 3 additionally include GP components with different kernels, as described in \Cref{sec:model_2,sec:model_3}.

For all models, the parameter posterior distributions are sampled with a Markov chain Monte Carlo Metropolis-Hasting algorithm as described by \citet{Sharma2017}. Each distribution is compiled from 50 independent chains of 200,000 iterations and a thinning factor of 10. Convergence is determined using the Gelman-Rubin statistic, with all chains having values under 1.02 \citep{Gelman1992}.

\subsubsection{Model 1: Keplerian-only\label{sec:model_1}}

For Model 1, we perform a Keplerian orbit fit to the RVs for the 9 parameters described above. No stellar activity mitigation scheme is adopted in this fiducial fit. Moreover, to search for longer-term accelerations, we also separately carry out two additional fits that model linear and and linear-plus-quadradic terms ($\dot{\gamma}$ and $\ddot{\gamma}$) for Model 1 only.

We impose uniform priors on all Keplerian model parameters. Based on the strong 4-day signal from the periodogram, we adopt $\mathcal{P}(P_b) = \mathcal{U}[0.5, 10.0]$~d and $\mathcal{P}(T_{0, b}) = \mathcal{U}[2458483.0, 2458487.5]$~d as priors for the orbital period and time of transit, respectively. We set the prior on $K_b$ to $\mathcal{P}(K_b) = \mathcal{U}[1.0, 1000.0]$~m s$^{-1}$. For the eccentricity parameters, we allow the full range of $\mathcal{P}(\sqrt{e_b} \sin \omega_*) = \mathcal{U}[-1, 1]$ and $\mathcal{P}(\sqrt{e_b} \cos \omega_*) = \mathcal{U}[-1, 1]$, which results in a uniform sampling of $\mathcal{P}(e_b) = \mathcal{U}[0, 1)$ and $\mathcal{P}(\omega_*) = \mathcal{U}[0, 2\pi]$, for $e_b$ and $\omega_*$ respectively. For the jitter terms, we adopt modified Jeffreys priors, $\mathcal{J}(1, 100)$~m s$^{-1}$, as defined by Equation 16 of \citet{Gregory2005},

\begin{equation}
    \mathcal{P}(\sigma;\sigma_a, \sigma_\mathrm{max}) = \frac{1}{\sigma_a + \sigma} \frac{1}{ \ln\left[ \left(\sigma_a + \sigma_\mathrm{max} \right) / \sigma_a \right] }.
\end{equation}

\noindent{}The modified Jeffreys prior behaves like a uniform prior for $\sigma < \sigma_a$ and a Jeffreys prior for $\sigma > \sigma_a$. The Jeffreys prior is scale invariant in that it treats each decade bin as having equal probability. For the offset terms, we use uniform priors bounded by the minimum and maximum of the RV measurements, $\mathcal{P}(\gamma_{\mathrm{HPF}_i}) = \mathcal{U}[\min(\mathrm{HPF}_i), \max(\mathrm{HPF}_i)]$. For the model fits that incorporate long-term trends, we adopt uniform priors of $\mathcal{P}(\dot{\gamma}) = \mathcal{U}[-1, 1]$~km s$^{-1}$ d$^{-1}$ and $\mathcal{P}(\ddot{\gamma}) = \mathcal{U}[-1, 1]$~km s$^{-1}$ d$^{-2}$.

We report the results of the Model 1 fit in \autoref{sec:phase_curve_and_posteriors}. \autoref{tab:model_params} summarizes the parameter constraints and \autoref{fig:model_1_results} displays the best-fit Keplerian signal, 1$\sigma$ and 3$\sigma$ confidence intervals, and associated joint posterior distributions in the form of a corner plot. We note that the model fits with linear and quadratic terms are consistent with zero acceleration and return nearly identical solutions. The resulting Bayesian Information Criterion \citep[BIC;][]{Schwarz1978, Raftery1986, Tran2022} for the Keplerian orbit-only, Keplerian orbit and acceleration, and Keplerian orbit, acceleration, and curvature models are $-12.0$, $-8.2$, and $-3.8$, respectively, which indicates that the data prefer the Keplerian orbit-only model. Thus, we report only the single-planet model without long-term acceleration or curvature terms.

The median and 1$\sigma$ posterior values for the Keplerian semi-amplitude and orbital period are $K_{b,\mathrm{Model\;1}} = 301 \pm 27$~m s$^{-1}$ and $P_{b,\mathrm{Model\;1}} = {3.986}_{-0.001}^{+0.001}$~d, respectively. Using our adopted stellar mass, $M_* = 0.69 \pm 0.07$~$M_\odot$ (\autoref{sec:mass+rad}), we find a minimum mass of $m_b \sin i_\mathrm{Model\;1} = 1.78_{-0.21}^{+0.22}$~$M_\mathrm{Jup}$ for HS Psc b. The best-fit eccentricity points to a modest value of $e_{b,\mathrm{Model\;1}} = 0.18$, although the posteriors are broad and consistent with circular orbits. The rms of the RV residuals after subtracting the best-fit Keplerian curve is 178 m s$^{-1}$.

The inferred posteriors of the RV jitter terms are $\sigma_{\mathrm{HPF_1,\; Model\;1}} = 161.3_{-24.6}^{+21.0}$~m s$^{-1}$ and $\sigma_{\mathrm{HPF_2,\; Model\;1}} = 17.4_{-17.4}^{+27.2}$~m s$^{-1}$. The large difference between $\sigma_{\mathrm{HPF_1,\; Model\;1}}$ and $\sigma_{\mathrm{HPF_2,\; Model\;1}}$ is likely driven by stronger stellar activity during the HPF$_1$ observations. The evolution of stellar activity over time is evident from the TESS light curves. The amplitude of photometric variability is approximately a factor of 2 greater in Sector 17 compared to Sector 57. As the HPF$_1$ and HPF$_2$ RVs are approximately coincident with TESS Sectors 17 and 57, respectively, this evolution could be driving differences between the jitter terms. Furthermore, the lower number and shorter time baseline of the HPF$_2$ RVs results in a diminished exploration of the stellar activity. This effect is amplified as a nightly cadence over this short observational window results in a limited sampling of the stellar rotation period with the HPF$_2$ observations. When phased to $P_\mathrm{rot} = 1.086$~d, the HPF$_2$ RVs cover only $\sim$1/3 of the stellar rotation phase space. This poor phase coverage can artificially dampen activity signals in the HPF$_2$ measurements, leading to a lower jitter estimate. Altogether these factors suggest that stellar activity is prominent in the RVs and necessitate a mitigation strategy.

\subsubsection{Model 2: Keplerian and Quasi-Periodic GP\label{sec:model_2}}

In Model 2, we simultaneously fit for a Keplerian signal and a GP model defined by a QP kernel. Under similar assumptions as for photometry (see \autoref{sec:rot_per}), the QP GP has been widely leveraged to model stellar activity in RVs \citep[e.g.,][]{Haywood2014, Grunblatt2015, Faria2016, Cloutier2017, Dai2017}. We adopt a GP model with a QP kernel as implemented in \texttt{pyaneti}:

\begin{equation} \label{eq:quasi_per_kernel_rv}
    k_\mathrm{QP}(t,t') = A^2 \: \mathrm{exp}\left( -\frac{(t - t')^2}{2\lambda_e^2} - \frac{\: \mathrm{sin}^2\left(\frac{\pi (t - t')}{P_\mathrm{GP}}\right)}{2\lambda_p^2} \right),%
\end{equation}

\noindent{}where $t$ and $t'$ represent the pairs of observations in time, and $A$, $\lambda_e$, $P_\mathrm{GP}$, and $\lambda_p$ are the same or analogous to $A$, $l$, $P_\mathrm{rot}$, and $\theta$ from \autoref{eq:quasi_per_kernel_lc}.

We adopt uniform priors on all Keplerian parameters. Based on the results from Model 1, we impose $\mathcal{P}(P_b) = \mathcal{U}[3.5, 4.5]$~d and $\mathcal{P}(T_{0,b} = \mathcal{U}[2458485.0, 2458488.0]$~d as priors for the orbital period and time of transit, respectively. All other priors for planetary parameters are the same as in Model 1.

For the QP kernel hyperparameters, we adopt $\mathcal{P}(A) = \mathcal{U}[0.0, 1000.0]$~m s$^{-1}$, $\mathcal{P}(\lambda_e) = \mathcal{U}[0.01, 5.0]$, $\mathcal{P}(\lambda_p) = \mathcal{U}[0.01, 5.0]$, and $\mathcal{P}(P_\mathrm{GP}) = \mathcal{N}(1.09, 0.02)$~d for $A$, $\lambda_e$, $\lambda_p$, and $P_\mathrm{GP}$, respectively. This last prior uses the MAP value and five times the standard deviation of $P_\mathrm{rot}$ measured from the TESS photometry. This is based on the assumption that the stellar activity signal is starspot-driven and will modulate at the stellar rotation period in both photometric and RV time series \citep[e.g.,][]{Aigrain2012, Tran2023}. The larger range allow the model to explore potential deviations in the rotation period across the large observational window, for instance from mid-latitude spots that could produce a spread of periodic signals as a result of differential rotation.

\autoref{tab:model_params_gp_qp} and \autoref{fig:model_2_results} in \autoref{sec:phase_curve_and_posteriors} summarizes and displays the results of the Model 2 fit, respectively. The inferred Keplerian semi-amplitude and orbital period from Model 2 are $K_{b,\mathrm{Model\;2}} = 264_{-86}^{+84}$~m s$^{-1}$ and $P_{b, \mathrm{Model\;2}} = 3.986_{-0.024}^{+0.005}$~d. This corresponds to a minimum mass of $m_b \sin i_\mathrm{Model\;2} = 1.46_{-0.43}^{+0.54}$~$M_\mathrm{Jup}$. The eccentricity is effectively unconstrained, as the posterior spans all possible values, although there is a slight preference for low eccentricities with $e_{b, \mathrm{Model\;2}} = 0.25_{-0.25}^{+0.19}$. The inferred RV jitter terms are $\sigma_{\mathrm{HPF_1,\; Model\;2}} = 8.0_{-8.0}^{+10.7}$~m s$^{-1}$ and $\sigma_{\mathrm{HPF_2,\; Model\;2}} = 8.1_{-8.1}^{+11.7}$~m s$^{-1}$. The rms of RV residuals for Model 2 is 108 m s$^{-1}$.

\begin{deluxetable*}{lccc}[!ht]
    \setlength{\tabcolsep}{16pt}
    \tablecaption{\label{tab:model_params_gp_m52} Parameter Priors and Posteriors from Keplerian and Mat\'ern--5/2 GP Fit (Model 3) to HPF RVs of HS Psc.}
    \tablehead{\colhead{Parameter} & 
    \colhead{Prior\tablenotemark{a}} & \colhead{MAP\tablenotemark{b}} & \colhead{Median $\pm$ 1$\sigma$}}
    \startdata
    \multicolumn{4}{c}{Fitted Parameters} \\
    \hline
    \multicolumn{4}{c}{Keplerian and Instrumental Parameters} \\
    $P_b$ (d) & $\mathcal{U}[3.5, 4.5]$ & $3.986$ & $3.986_{-0.003}^{+0.044}$ \\
    $T_{0, b}$ (d) & $\mathcal{U}[2458485.0, 2458488.0]$ & $2458486.565$ & $2458486.422_{-0.523}^{+0.429}$ \\
    $K_b$ (m s$^{-1}$) & $\mathcal{U}[1.0, 1000.0]$ & $281.6$ & $268_{-92}^{+91}$ \\
    $\sqrt{e_b} \sin{\omega_*}$ & $\mathcal{U}[-1, 1]$ & $0.10$ & $0.01_{-0.50}^{+0.45}$ \\
    $\sqrt{e_b} \cos{\omega_*}$ & $\mathcal{U}[-1, 1]$ & $-0.38$ & $-0.10_{-0.39}^{+0.38}$ \\
    $\sigma_{\mathrm{HPF}_{1}}$ (m s$^{-1}$) & $\mathcal{J}(1, 100)$ & $0.1$ & $8.2_{-8.2}^{+11.2}$ \\
    $\sigma_{\mathrm{HPF}_{2}}$ (m s$^{-1}$) & $\mathcal{J}(1, 100)$ & $0.4$ & $8.2_{-8.2}^{+11.6}$ \\
    $\gamma_{\mathrm{HPF}_{1}}$ (km s$^{-1}$) & $\mathcal{U}[-1.0922, 0.9655]$ & $-0.025$ & $-0.037_{-0.051}^{+0.047}$ \\
    $\gamma_{\mathrm{HPF}_{2}}$ (km s$^{-1}$) & $\mathcal{U}[-0.8858 , 1.0037]$ & $0.076$ & $0.068_{-0.088}^{+0.090}$ \\
    \multicolumn{4}{c}{Mat\'ern--5/2 GP Kernel Hyperparameters} \\
    $A$ (m s$^{-1}$) & $\mathcal{U}[0.0, 1000.0]$ & $139.5$ & $187.2_{-55.6}^{+41.1}$ \\
    $\lambda$ & $\mathcal{U}[0.001, 10.0]$ & $0.02$ & $0.09_{-0.08}^{+0.04}$ \\
    \hline
    \multicolumn{4}{c}{Derived Parameters} \\
    \hline
    $m_b \sin i$\tablenotemark{c} $(M_\mathrm{Jup})$ & $\cdots$ & $1.71$ & $1.46_{-0.44}^{+0.56}$ \\
    $a_b$ (AU) & $\cdots$ & $0.0436$ & $0.0435_{-0.0017}^{+0.0017}$ \\
    $T_\mathrm{peri, b}$ (d) & $\cdots$ & $2458487.21$ & $2458486.71_{-0.91}^{+1.46}$ \\
    $e_b$ & $\cdots$ & $0.16$ & $0.27_{-0.27}^{+0.21}$ \\
    $\omega_*$ $(\degree)$ & $\cdots$ & $165.30$ & $176.25_{-102.74}^{+101.88}$ \\
    \enddata
    \tablenotetext{a}{$\mathcal{U}[a, b]$ refers to the uniform distribution bounded by $a$ and $b$. $\mathcal{J}(a, b)$ refers to the modified Jeffreys prior as defined in Equation 6 of  \citet{Gregory2005}, $\mathcal{P}(x) = 1/(a + x) \cdot 1/\ln \left[\left(a + b\right) / a\right]$.}
    \tablenotetext{b}{MAP refers to the \textit{maximum a posteriori} value.}
    \tablenotetext{c}{Planetary mass is derived assuming a stellar mass of $M_* = 0.69 \pm 0.07$ $M_\odot$.}
\end{deluxetable*}

\begin{figure*}[!ht]
        \centering
        \includegraphics[width=0.89\linewidth]{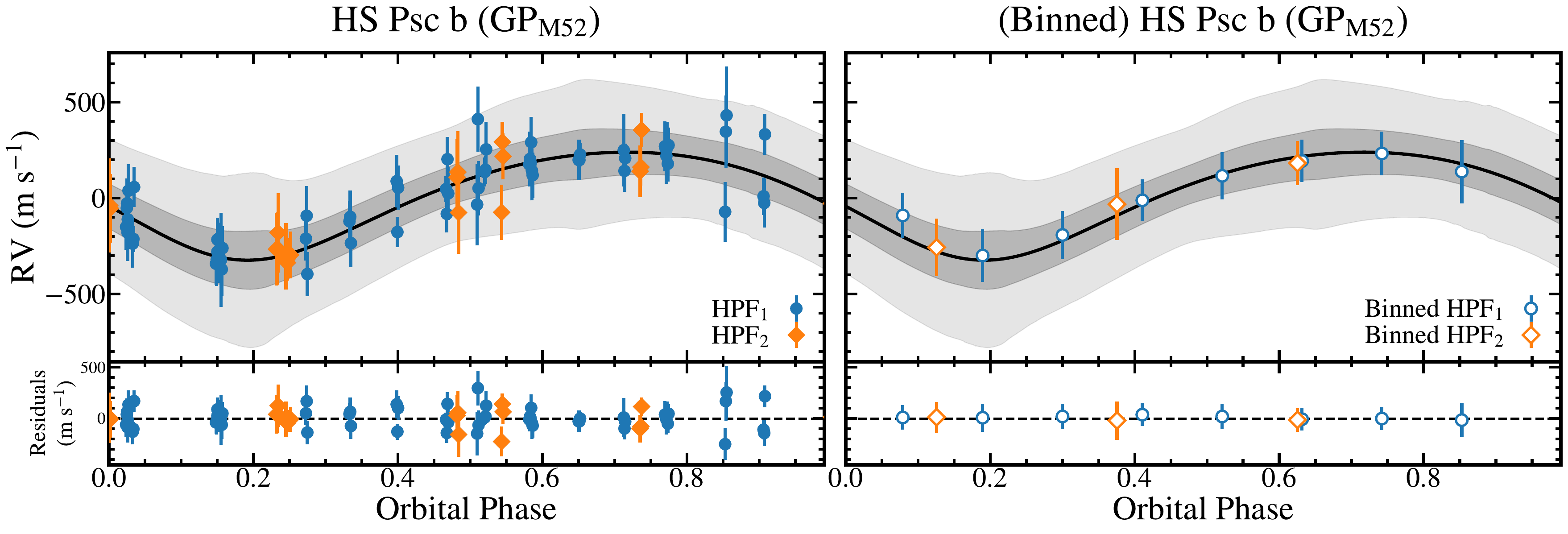}
        
        \hspace{-3mm} \includegraphics[width=0.925\linewidth]{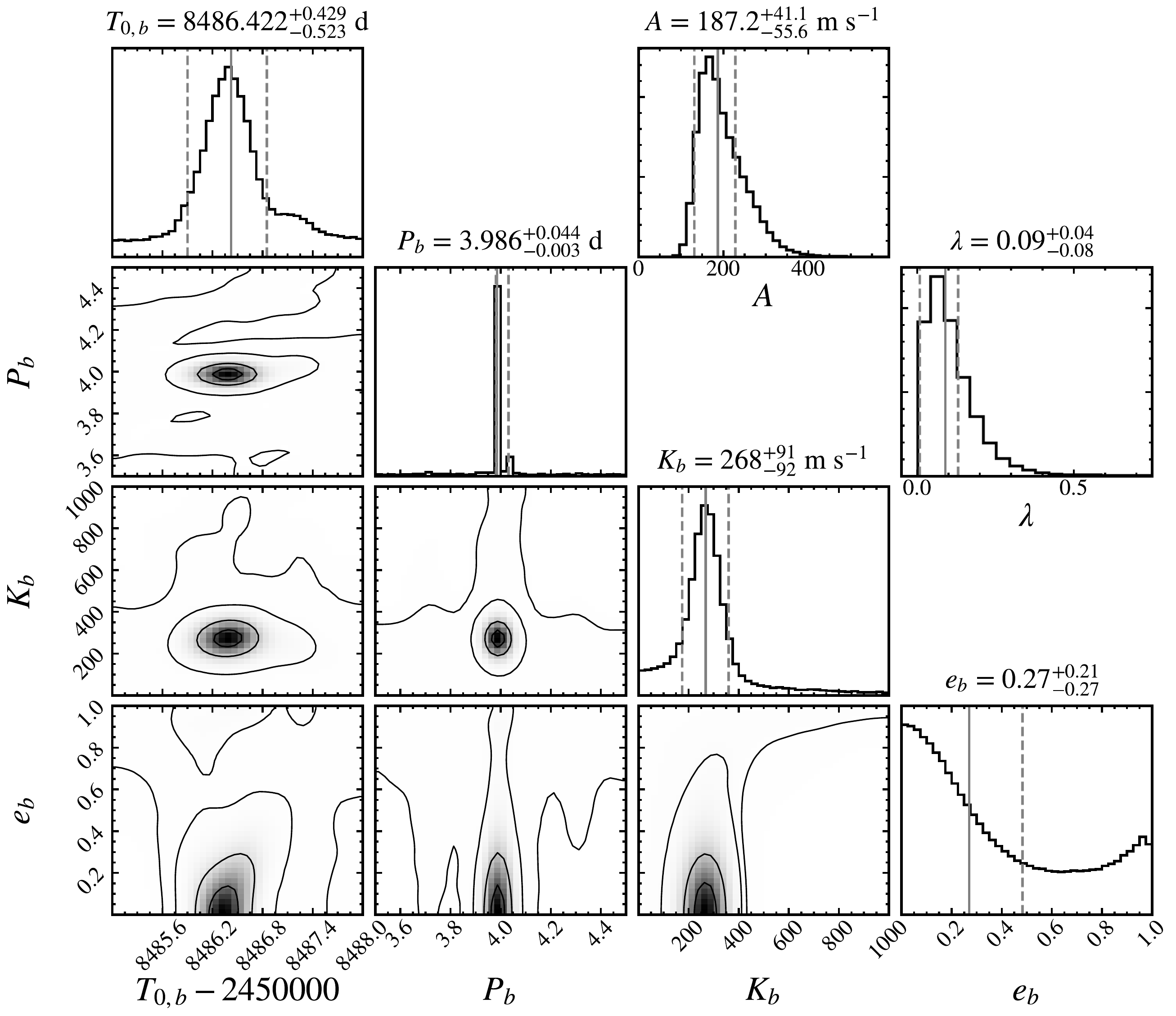}
	    \caption{Top: RV curve of HS Psc b phase folded to the best-fit orbital period from Model 3. The blue and orange points denote the two different observing seasons, HPF$_1$ and HPF$_2$, respectively, and the best-fit RV model is shown as the solid black line. The fit residuals are plotted in the lower panels. The colored error bars are nominal RV errors and the grey error bars include the systematic jitter terms. 1$\sigma$ and 3$\sigma$ confidence intervals are plotted as grey shaded regions. The right panel displays the same best-fit orbit with the phased RV points representing median values in bins of $\approx$0.1 and $\approx$0.2 of the phase for the HPF$_1$ and HPF$_2$ RVs, respectively. The bottom panels display the joint posterior distributions of $T_{0, b}$, $P_b$, $K_b$, and $e_b$ from the Model 3 fit of HS Psc b. The diagonal panels show the marginalized distribution for each parameter. The upper right panels show the marginalized distributions for the M$\sfrac{5}{2}$ kernel parameters.}
	\label{fig:model_3_results}
\end{figure*}

\subsubsection{Model 3: Keplerian and Mat\'ern--5/2 GP\label{sec:model_3}} 

For Model 3, we perform a similar joint Keplerian-plus-GP model fit as Model 2, but with a GP defined by a Mat\'ern--5/2 (M$\sfrac{5}{2}$) kernel instead of a QP kernel. Our goal is to assess how changes to the GP stellar activity model can impact the inferred planetary parameters. Furthermore, this change allows us to account for the possibility that the stellar activity signals manifest differently in the RVs as compared to the TESS photometry. The Mat\'ern family of kernels are also widely adopted in GP regression to model more stochastic behavior. GP models incorporating the M$\sfrac{5}{2}$ kernel and its derivative have been found to reasonably match stellar activity in solar RV data \citep{Gilbertson2020}.

We apply a GP model with a M$\sfrac{5}{2}$ kernel as defined in \texttt{pyaneti}:

\begin{equation} \label{eq:matern_52_kernel_rv}
    k_\mathrm{M\sfrac{5}{2}}(t,t') = A^2 \left( 1 + t_{\sfrac{5}{2}} + \frac{t^{2}_{\sfrac{5}{2}}}{3} \right) \mathrm{exp} \left( -t_{\sfrac{5}{2}} \right),
\end{equation}

\noindent{}with $t_{\sfrac{5}{2}} \equiv \sqrt{5}\left|t - t'\right| \lambda^{-1}$, where $\lambda$ is the length of local variations.

We adopt uniform priors for all parameters in the joint Model 3 fit. For the planetary parameters, we impose the same priors as in Model 2. For the M$\sfrac{5}{2}$ kernel hyperparameters, we choose $\mathcal{P}(A) = \mathcal{U}[0.0, 1000.0]$~m s$^{-1}$ for $A$ and $\mathcal{P}(\lambda) = \mathcal{U}[0.001, 10.0]$ for $\lambda$.

\autoref{tab:model_params_gp_m52} details the results of the Model 3 fit. \autoref{fig:model_3_results} displays the best-fit phased RV curves and posterior distributions of the planetary and M$\sfrac{5}{2}$ kernel parameters. For Model 3, the inferred Keplerian semi-amplitude is $K_{b,\mathrm{Model\;3}} = 268_{-92}^{+91}$~m s$^{-1}$, which corresponds to a minimum planetary mass of $m_b \sin i_\mathrm{Model\;3} = 1.46_{-0.44}^{+0.56}$~$M_\mathrm{Jup}$. The orbital period is $P_{b, \mathrm{Model\;3}} = 3.986_{-0.003}^{+0.044}$~d. Similar to Model 2, the eccentricity is unconstrained with some preference for low values $(e_{b, \mathrm{Model\;3}} = 0.27_{-0.27}^{+0.21})$ and the RV jitter terms are $\sigma_{\mathrm{HPF_1,\; Model\;3}} = 8.2_{-8.2}^{+11.2}$~m s$^{-1}$ and $\sigma_{\mathrm{HPF_2,\; Model\;3}} = 8.2_{-8.2}^{+11.6}$~m s$^{-1}$. The rms of the Model 3 RV residuals is 99~m s$^{-1}$.

\begin{deluxetable*}{cccccccc}[!t]
    \setlength{\tabcolsep}{8pt}
    \tablecaption{Metrics for best-fit model comparison.} \label{tab:model_criterion}
    \tablehead{\colhead{Model} & \colhead{ln $\mathcal{L}$} & \colhead{$k$} & \colhead{$N$} & \colhead{BIC} & \colhead{AIC} & \colhead{AIC\textsubscript{c}} & \colhead{Akaike Weight}}
    \startdata
    Model 1 (Keplerian-only) & $25.92$ & $9$ & $83$ & $-12.07$ & $-33.84$ & $-31.37$ & $<0.01$ \\
    Model 2 (Keplerian + QP GP) & $35.48$ & $13$ & $83$ & $-13.51$ & $-44.95$ & $-39.68$ & $0.08$ \\
    Model 3 (Keplerian + M$\sfrac{5}{2}$ GP) & $\mathbf{35.09}$ & $\mathbf{11}$ & $\mathbf{83}$ & $\mathbf{-21.57}$ & $\mathbf{-48.18}$ & $\mathbf{-44.46}$ & $\mathbf{0.92}$ \\
    \enddata
    \tablecomments{Lower metric values indicate a preferred model. The bolded model (Keplerian-plus-M$\sfrac{5}{2}$ GP) is adopted as the overall best fit solution.}
\end{deluxetable*}

\begin{figure*}[!th]
    \centering
    \includegraphics[width=1.0\linewidth]{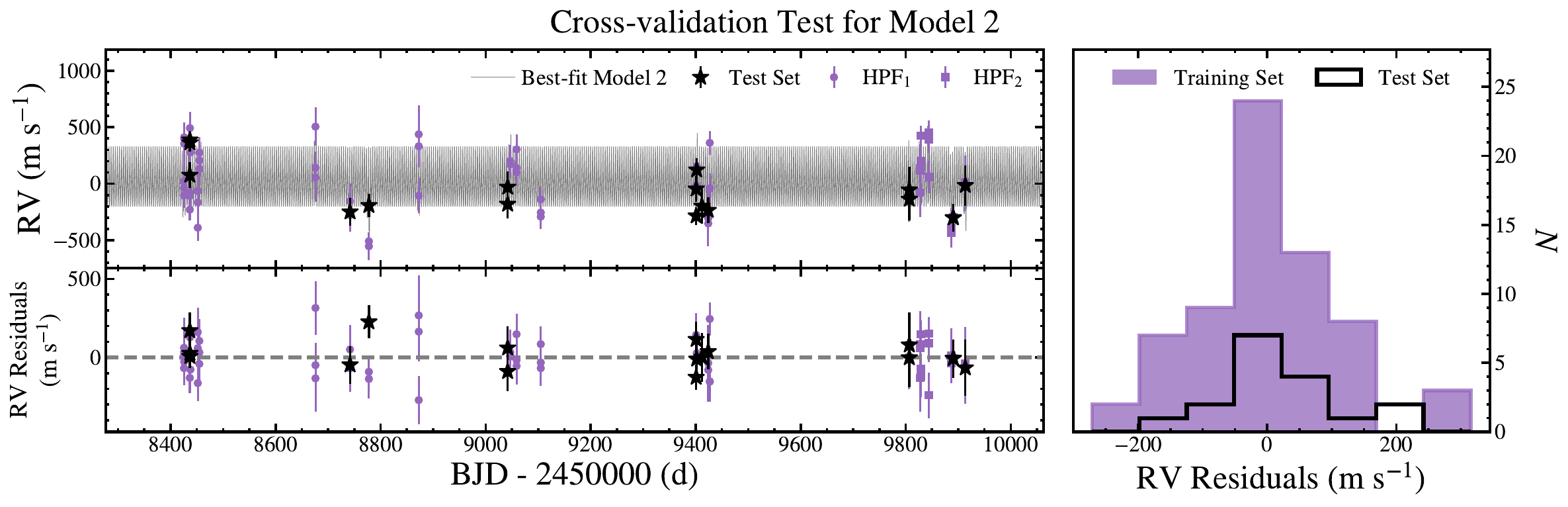}
    \includegraphics[width=1.0\linewidth]{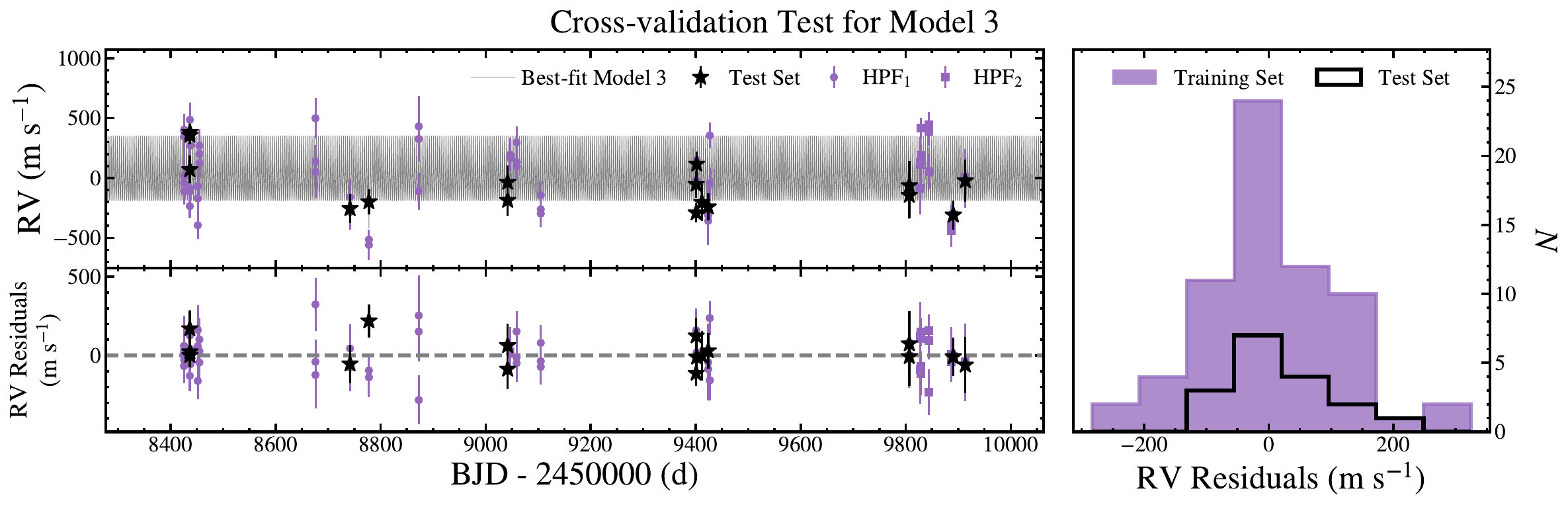}
    \caption{Cross-validation tests for Model 2 (top) and Model 3 (bottom). Left: Best-fit joint Keplerian and GP models are plotted as solid black lines. The training and test RV data sets are displayed as purple points and black stars, respectively. The residuals between the models and both datasets are displayed on the bottom. Right: The distribution of RV residuals for the training and test data sets are plotted in purple and black, respectively. The range and structure of the residuals are consistent between the training and test sets, supporting the predictive power of the best-fit models and providing evidence against overfitting.}
    \label{fig:cross_valid_gp}
\end{figure*}

\subsubsection{Model Comparison, GP Cross-Validation, and Adopted Parameters}
All inferred posteriors for planetary and instrumental parameters are consistent among the three models within $1\sigma$ except for the HPF$_1$ RV jitter term. The posterior values of $\sigma_{\mathrm{HPF_1}}$ are similar for both Models 2 and 3, which are reduced by a factor of approximately 20 compared to the inferred Model 1 values. This suggests that the higher Model 1 $\sigma_{\mathrm{HPF_1}}$ values are driven by correlated stellar activity signals as both GP models are able to remove this additional scatter. Furthermore, Models 2 and 3, which employ GPs with different kernel choices, yield nearly identical planetary parameters. This provides evidence that the GP models are identifying similar correlated signals and are robust against kernel selection, which can affect the inferred planetary properties \citep[e.g.,][]{Benatti2021}.

The GP models also decrease the rms of the RV residuals. The residual rms values are 178, 109, and 99 m s$^{-1}$ for Models 1, 2, and 3, respectively. Altogether the lower overall residuals, the reduced $\sigma_{\mathrm{HPF_1}}$ parameter, and the consistent planetary values suggest that the GP models are effectively mitigating stellar activity contributions to the RVs of HS Psc.

Finally, the uncertainties in the RV semi-amplitudes and eccentricities for Models 2 and 3 are larger than for Model 1. The 1$\sigma$ uncertainties of Models 2 and 3 increase by a factor of approximately 3 for $K_b$ and 2 for $e_b$ as compared to Model 1. The lower Model 1 uncertainties can be attributed to the large planetary amplitude, long time baseline, and high number of RV observations. However, the posterior width may not fully capture the effects of stellar activity, which are not explicitly built into the model. These larger uncertainties present in Models 2 and 3 better reflect the difficulty of detecting and appropriately characterizing a young planet candidate such as HS Psc b.

Altogether we find that a joint GP model better predicts the HPF RVs compared to a Keplerian-only model. However, GP regression can be susceptible to overfitting \citep[e.g.,][]{Aigrain2022, Blunt2023}, leading to systematic biases that affect the interpretation of a planetary signal. This is particularly true for young systems where stellar activity signals are large.

To assess the possibility of Models 2 and 3 overfitting the data, we apply a cross-validation, or ``train-test-split'', test as described in \citet{Blunt2023}. Cross-validation is a procedure designed to test the predictive performance of a model \citep{Gelfand1992, Aigrain2022}. As implemented in \citet{Blunt2023}, a validation sample is first constructed by splitting the data into training and test sets, which are comprised of 80\% and 20\% of the data, respectively. The model is then conditioned on the training set and the best-fit model predictions at the times of the test set are compared to the test set values. If the predictions produce similar residuals for the withheld data as with the conditioned data, then the model is predictive and can be considered robust against overfitting.

We randomly withhold 17 RVs out of the 83 observations (amounting to 20\% of the data) and conduct this cross-validation test for both Models 2 and 3. \autoref{fig:cross_valid_gp} displays the training and test data sets, best-fit model predictions, and fit residuals from this cross-validation test for Models 2 and 3. For both models, the residuals of the test set have consistent structure and spread with the training set residuals. This test is repeated twice more with different data split variations. The results in all three instances are consistent. We interpret the similarity between the residuals of the training and test sets as evidence supporting the predictive nature of the best-fit models, indicating that Models 2 and 3 do not overfit the RV data.

To determine which model is favored by the data, we further evaluate comparison metrics between Models 1, 2, and 3. \autoref{tab:model_criterion} reports different model selection criteria for each of the three models, including the BIC, both the uncorrected and corrected Akaike Information Criterion \citep[AIC and AIC$_c$;][]{Akaike1998, Tran2022} and the Akaike weights \citep{Akaike1981, Burnham2004, Liddle2007}. The BIC comparison suggests that there is strong evidence in favor of Model 3 ($\Delta\text{BIC} > 5$ against Models 1 and 2). The Akaike weights, or relative likelihoods, also prefer Model 3 over Models 1 and 2 (92\% favorability).

The consistent inferred parameters, reduced residual rms, cross-validation tests, and model selection criteria all indicate that Model 3 performs the best at robustly detecting the Keplerian signal and modeling correlated stellar activity signals. As a result, we adopt the results from Model 3, the joint Keplerian and M$\sfrac{5}{2}$ GP model, as the most appropriate values for HS Psc b.

\section{Discussion\label{sec:discussion}}

\subsection{Exclusion of a Low-Inclination Brown Dwarf or Stellar Companion}

RV observations alone cannot measure the true mass of a Keplerian signal. RV surveys are thus susceptible to false-positive scenarios in which a BD or stellar companion on a face-on orbit masquerades as a planet \citep[e.g.,][]{Wright2013}. In \autoref{sec:rv_fit}, we report a minimum mass of $m \sin i_* = 1.46$~$M_\mathrm{Jup}$. For HS Psc b to have a true mass larger than the BD limit ($\geq 13$~$M_\mathrm{Jup}$), its orbital inclination must be $i \leq 6.4\degr$. Assuming an isotropic distribution for inclinations, the \textit{a priori} probability of this scenario is $P \approx 0.63\%$. For a stellar mass companion ($\geq 75$ $M_\mathrm{Jup}$), this probability decreases to $P \approx 0.019\%$. However, these probabilities only reflect the chances of a false positive planet signal for this particular star and do not take into account our broader survey size; the larger the survey, the more false positive planets should be found. Binomial statistics can be used to correct for this sample size by determining the probability of a success (planet) given an underlying rate (the false positive probability) and a sample size (104 stars, in this case). This result will assume that each star in the sample has a short-period stellar or substellar companion, so the resulting probability needs to be multiplied by the actual occurrence rate of these companions.

Combining this probability of an inclined BD, binomial statistics based on our survey, and a conservative estimate of 1\% for the occurrence rate of BDs, we calculate the probability of observing at least 1 false-positive event in our sample of 104 systems from this system as $P_\mathrm{FP} = 0.48\%$. This false-positive rate is an upper limit as the occurrence rate of close-in ($a \lesssim 5$~au) BDs is $\lesssim1\%$ \citep[e.g.,][]{Grether2006, Sahlmann2011, Santerne2016} and even lower at closer separations \citep[$P \lesssim 100$~d;][]{Ma2014, Ranc2015, Csizmadia2015, Kiefer2019}. The true probability of a false alarm scenario is therefore likely to be $<0.48\%$. Moreover, we note that the stellar inclination is likely high (\autoref{sec:rot_per}) and hot Jupiters around cool stars usually have low obliquities \citep[e.g.,][]{Winn2010, Albrecht2022}, so it would be unusual for the planet inclination to differ substantially from $90\degr$. Based on these estimates, we confidently exclude the possibility that HS Psc b is a BD or star on an inclined orbit.

\subsection{Timescales for Tidal Circularization and High-eccentricity Migration}

As a young hot Jupiter, HS Psc b could offer rare constraints on the timescales of different migration processes such as tidal circularization and high-eccentricity migration. Tidal interactions between close-in planets and their host stars lead to an exchange of angular momentum and, as a result, evolution in orbital separation and eccentricity. The characteristic timescale for this mechanism is typically short compared to the characteristic age of several Gyr for field stars. A general form for the tidal circularization timescale is given by Equation 3 of \citet{Adams2006},

\begin{align}
\begin{split} \label{eq:tidal_circ}
    \tau_\mathrm{cir} &\approx 1.6 \; \mathrm{Gyr} \; \left( \frac{Q_p}{10^6} \right) \left( \frac{m_p}{M_\mathrm{Jup}} \right) \\ 
    &\times \left( \frac{M_*}{M_\odot} \right)^{-3/2} \left( \frac{r_p}{R_\mathrm{Jup}} \right)^{-5} \left( \frac{a}{0.05 \; \mathrm{AU}} \right)^{13/2},
\end{split}
\end{align}

\noindent{}where $Q_P$ is the tidal quality factor, generally $10^5$--$10^6$, $m_p$ and $r_p$ are the planetary mass and radius, respectively, $M_*$ is the stellar host mass, and $a$ is the orbital separation. This relation holds for systems where the planetary orbital period is greater than the stellar rotational period \citep{Goldreich1966, Jackson2008}. However, HS Psc b operates in the super-synchronous rotation regime, where $P_\mathrm{orb} > P_\mathrm{rot}$ \citep{Ferraz-Mello2008}. In this state, the assumptions in \autoref{eq:tidal_circ} break down and tidal forces operate less efficiently. As a result, circularization timescales estimated by \autoref{eq:tidal_circ} should be treated as lower limits \citep{Jackson2008}.

Coupled with the age of the system, this lower limit can provide us with a sense of the minimum time it should take for a planet like HS Psc b to tidally circularize if it had a high eccentricity after migrating to its current orbital distance. Adopting the median values reported in this work and assuming radii of \{1.0, 1.25, 1.5\} $R_\mathrm{Jup}$,\footnote{Giant plants are inflated at young ages, then cool and contract over time. Based on the mass (1--2 $M_\mathrm{Jup}$) and age ($\approx$100 Myr) of HS Psc b, hot-start giant planet evolutionary models predict a radius of $\sim$1.2 $R_\mathrm{Jup}$ \citep[e.g.,][]{Burrows2001, Baraffe2003, Morley2024}.} the ranges of tidal circularization timescales of HS Psc b are $\tau_\mathrm{cir} =$ \{0.18--1.79, 0.09--0.89, 0.05--0.48\} Gyr.

These estimates suggest the earliest timescale for tidal circularization is on the order of several tens of Myr. This means that if the current eccentricity of HS Psc b is zero, then if it migrated recently through high-eccentricity migration, the earliest that process could have occurred was a few tens of Myr ago. It could also have migrated this way soon after formation, setting an upper limit to this process of $\approx$130 Myr. A low eccentricity is also consistent with disk migration early on. On the other hand, if HS Psc b currently has a modest or high eccentricity, then it should still be in the process of circularizing and high-eccentricity migration could have occurred at any point in the planet's history.

We stress that while the eccentricity posteriors of the adopted model (Model 3) prefer lower values, there is power across all eccentricities and this parameter is not well constrained. Follow-up RV observations are needed to more confidently rule out higher eccentricities and better constrain the evolutionary history and migration timescale of HS Psc b. Irrespective of the specifc timing or channel, the discovery of HS Psc b implies that HJs can form or migrate to their current locations by 130~Myr.

\subsection{Future Observations}

Optical RV observations can further validate and characterize HS Psc b. Modulations arising from the presence of starspots are wavelength dependent; RV amplitudes of starspot-driven variability have been shown to be higher in the optical than in the NIR by a factor of $\approx$2 for systems at this age \citep[e.g.,][]{Prato2008, Mahmud2011, Crockett2012, Bailey2012, Gagne2016, Tran2021}. Confirming that the RV amplitude, orbital period, and orbital phase of optical RVs are consistent with the NIR HPF observations would further support the planetary nature of HS Psc b. Additional RVs can also be used to search for other companions, such as smaller closer-in or more distant giant planets, to further inform the formation and migration of the system. Moreover, more precise optical RVs can refine the eccentricity of HS Psc b and establish whether it is actively undergoing tidal circularization.

\section{Summary\label{sec:summary}}

We have presented the discovery and characterization of a young hot Jupiter candidate orbiting a member of the $133_{-20}^{+15}$~Myr AB Dor moving group as part of the Epoch of Giant Planet Migration planet search program. Below, we summarize our main conclusions:

\begin{itemize}
    \item We obtained 83 NIR RVs of HS Psc with HPF at the HET. Using \texttt{SERVAL}-based least-squares matching and \texttt{SpecMatch-Emp}-based empirical spectral matching algorithms, we extract the relative RVs, dLWs, CRXs, and the \ion{Ca}{2} infrared triplet line indices.
    
    \item We derive an effective temperature of $T_\mathrm{eff} = 4203 \pm 116$~K, a metallicity of $\mathrm{[Fe/H]} = -0.05 \pm 0.09$~dex, a surface gravity of $\log g = 4.66 \pm 0.03$~dex, and a stellar projected rotational velocity of $v \sin i_* = 29.7 \pm 3.1$~km s$^{-1}$ for HS Psc. We infer a stellar mass of $M_* = 0.69 \pm 0.07$~$M_\odot$ and radius of $R_* = 0.65 \pm 0.07$~$R_\odot$ for HS Psc using the \texttt{ARIADNE} and \texttt{isochrones} packages. These values are consistent with its spectral type of K7V, stellar age, and previous literature values.
    
    \item Our HPF RVs over 4 years reveal a periodicity at $P = 3.99$~d. This period is not commensurate with an integer harmonic of the stellar rotation period measured from TESS photometry ($P_\mathrm{rot} = 1.086 \pm 0.003$~d). Furthermore, these RV measurements do not correlate significantly with associated activity indicators, supporting a planetary origin for the observed signal.

    \item A joint Keplerian and M$\sfrac{5}{2}$ GP stellar activity model fit to the HPF RVs yields a minimum mass of $m_b \sin i = 1.46_{-0.44}^{+0.57}$~$M_\mathrm{Jup}$, an orbital period of $P_b = 3.986_{-0.003}^{+0.044}$~d, and a broad eccentricity constraint with a slight preference for low values. No evidence of a longer-term acceleration is evident. HS Psc b is unlikely to be a BD or star on a face-on orbit.

    \item As a young, close-in giant planet, HS Psc b may have undergone high-eccentricity tidal migration. If so, we estimate a lower limit of several tens of Myr for the tidal circularization timescale of HS Psc b. The age of HS Psc places an upper limit on the migration timescale of $\approx$130~Myr. Disk migration is also possible if HS Psc b has a low eccentricity. A modest or high eccentricity would imply that it is still undergoing circularization. Additional high-precision RV observations will help confirm HS Psc b, refine the orbit and minimum mass, and constrain its orbital evolutionary history.
    
\end{itemize}

HS Psc b joins only a small handful of other hot Jupiter candidates that have both robust age constraints and planetary (minimum) mass measurements. HS Psc b is an excellent target for future observations with precision optical and IR spectrographs. If confirmed with follow-up radial velocities, HS Psc b will be one of the youngest hot Jupiters discovered to date.

\section{Acknowledgements}

The authors would like to thank Marvin Morgan, Kyle Franson, and Adam Kraus for insightful discussions on the high-eccentricity tidal migration, orbital circularization, and thermal evolution of close-in giant planets. The authors would also like to thank Chad Bender, Steven Janowiecki, Greg Zeimann, the HPF team, and all the resident astronomers and telescope operators at the HET for supporting these observations and data processing. The authors are grateful to the referee for their helpful comments, which improved the quality of this manuscript.

Q.H.T. and B.P.B. acknowledge the support from a NASA FINESST grant (80NSSC20K1554). B.P.B. acknowledges support from the National Science Foundation grant AST-1909209, NASA Exoplanet Research Program grant 20-XRP20$\_$2-0119, and the Alfred P. Sloan Foundation. GS acknowledges support provided by NASA through the NASA Hubble Fellowship grant HST-HF2-51519.001-A awarded by the Space Telescope Science Institute, which is operated by the Association of Universities for Research in Astronomy, Inc., for NASA, under contract NAS5-26555.

These results are based on observations obtained with the Habitable-zone Planet Finder Spectrograph on the HET. The HPF team acknowledges support from NSF grants AST-1006676, AST-1126413, AST-1310885, AST-1517592, AST-1310875, ATI 2009889, ATI-2009982, AST-2108512, and the NASA Astrobiology Institute (NNA09DA76A) in the pursuit of precision radial velocities in the NIR. The HPF team also acknowledges support from the Heising-Simons Foundation via grant 2017-0494. The Hobby-Eberly Telescope is a joint project of the University of Texas at Austin, the Pennsylvania State University, Ludwig-Maximilians-Universität München, and Georg-August Universität Gottingen. The HET is named in honor of its principal benefactors, William P. Hobby and Robert E. Eberly. We acknowledge the Texas Advanced Computing Center (TACC) at The University of Texas at Austin for providing high performance computing, visualization, and storage resources that have contributed to the results reported within this paper. Computations for this research were also performed on the Pennsylvania State University’s Institute for Computational and Data Sciences Advanced CyberInfrastructure (ICDS-ACI, now known as Roar), including the CyberLAMP cluster supported by NSF grant MRI1626251.

We would like to acknowledge that the HET is built on Indigenous land. Moreover, we would like to acknowledge and pay our respects to the Carrizo \& Comecrudo, Coahuiltecan, Caddo, Tonkawa, Comanche, Lipan Apache, Alabama-Coushatta, Kickapoo, Tigua Pueblo, and all the American Indian and Indigenous Peoples and communities who have been or have become a part of these lands and territories in Texas, here on Turtle Island.

This paper includes data collected by the TESS mission. Funding for the TESS mission is provided by the NASA's Science Mission Directorate. This work presents results from the European Space Agency (ESA) space mission Gaia. Gaia data are being processed by the Gaia Data Processing and Analysis Consortium (DPAC). Funding for the DPAC is provided by national institutions, in particular the institutions participating in the Gaia MultiLateral Agreement (MLA).

This research has made use of the VizieR catalogue access tool, CDS, Strasbourg, France (DOI: 10.26093/cds/vizier). The original description of the VizieR service was published in 2000, A\&AS 143, 23. This publication makes use of data products from the Two Micron All Sky Survey, which is a joint project of the University of Massachusetts and the Infrared Processing and Analysis Center/California Institute of Technology, funded by the National Aeronautics and Space Administration and the National Science Foundation. This publication makes use of data products from the Wide-field Infrared Survey Explorer, which is a joint project of the University of California, Los Angeles, and the Jet Propulsion Laboratory/California Institute of Technology, funded by the National Aeronautics and Space Administration.

\facilities{HET (HPF), TESS}

\software{\texttt{ARIADNE} \citep{Vines2022},
    \texttt{astropy} \citep{Astropy2013, Astropy2018, Astropy2022},
    \texttt{astroquery} \citep{Ginsburg2019},
    \texttt{barycorrpy} \citep{Kanodia2018a},
    \texttt{celerite2} \citep{Foreman-Mackey2017, Foreman-Mackey2018},
    \texttt{dustmaps} \citep{Green2018},
    \texttt{dynesty} \citep{Higson2019, Speagle2020, Koposov2023},
    \texttt{HxRGproc} \citep{Ninan2018},
    \texttt{isochrones} \citep{Morton2015},
    \texttt{lightkurve} \citep{Lightkurve2018},
    \texttt{matplotlib} \citep{Hunter4160265},
    \texttt{MultiNest} \citep{Feroz2009, Feroz2019},
    \texttt{numpy} \citep{vanderWalt2011},
    \texttt{pandas} \citep{mckinney-proc-scipy-2010},
    \texttt{pyaenti} \citep{Barragan2019, Barragan2022},
    \texttt{PyMultiNest} \citep{Buchner2014},
    \texttt{scipy} \citep{Virtanen2020},
    \texttt{SERVAL} \citep{Zechmeister2018}.
    }

\appendix
\restartappendixnumbering

\section{HPF RV and Activity Indicators\label{sec:appendix_measurements}}

\autoref{tab:hpf_measurements} lists the measured relative HPF RVs and associated stellar activity indicators of HS Psc b. See \autoref{sec:hpf} for details.

\startlongtable
    \begin{deluxetable*}{cccccccc}
    \setlength{\tabcolsep}{4pt}
    \tablecaption{Relative HPF RVs, activity indicators (dLW and CRX), and line indices for the \ion{Ca}{2} IRT lines measurements and uncertainties \label{tab:hpf_measurements}}
    \tablehead{\colhead{BJD\textsubscript{TDB}} & \colhead{RV} & \colhead{dLW} & \colhead{CRX} & \colhead{\ion{Ca}{2} IRT 1} & \colhead{\ion{Ca}{2} IRT 2} & \colhead{\ion{Ca}{2} IRT 3} & \colhead{Instrument} \\
    \colhead{(d)} & \colhead{(m s$^{-1}$)} & \colhead{(m$^2$ s$^{-2}$)} & \colhead{(m s$^{-1}$ Np$^{-1}$)} & \colhead{} & \colhead{} & \colhead{} & \colhead{}}
    \startdata
    2458425.6292 & -17.5 $\pm$ 190.0 & 3346.0 $\pm$ 1061.7 & 183.4 $\pm$ 234.8 & 0.919 $\pm$ 0.005 & 0.866 $\pm$ 0.006 & 0.820 $\pm$ 0.005 & HPF$_1$ \\
    2458425.6333 & -144.2 $\pm$ 108.2 & 1691.0 $\pm$ 708.1 & 100.1 $\pm$ 131.0 & 0.882 $\pm$ 0.005 & 0.854 $\pm$ 0.007 & 0.788 $\pm$ 0.005 & HPF$_1$ \\
    2458425.6371 & -76.3 $\pm$ 75.2 & 1424.3 $\pm$ 973.8 & 1.2 $\pm$ 95.3 & 0.875 $\pm$ 0.005 & 0.788 $\pm$ 0.007 & 0.771 $\pm$ 0.006 & HPF$_1$ \\
    2458425.8566 & 371.1 $\pm$ 130.4 & 251.4 $\pm$ 646.0 & -103.6 $\pm$ 161.5 & 0.907 $\pm$ 0.004 & 0.859 $\pm$ 0.005 & 0.794 $\pm$ 0.004 & HPF$_1$ \\
    2458425.8605 & 358.5 $\pm$ 73.1 & 1613.4 $\pm$ 592.2 & -2.7 $\pm$ 93.6 & 0.875 $\pm$ 0.005 & 0.805 $\pm$ 0.007 & 0.771 $\pm$ 0.005 & HPF$_1$ \\
    2458425.8644 & 313.5 $\pm$ 146.2 & -379.1 $\pm$ 762.8 & -112.3 $\pm$ 182.3 & 0.882 $\pm$ 0.004 & 0.840 $\pm$ 0.005 & 0.779 $\pm$ 0.004 & HPF$_1$ \\
    2458436.6067 & -146.1 $\pm$ 67.9 & -791.0 $\pm$ 722.2 & 18.4 $\pm$ 86.0 & 0.875 $\pm$ 0.005 & 0.845 $\pm$ 0.006 & 0.774 $\pm$ 0.005 & HPF$_1$ \\
    2458436.6107 & -268.5 $\pm$ 97.6 & -697.6 $\pm$ 751.3 & -159.5 $\pm$ 104.2 & 0.899 $\pm$ 0.004 & 0.870 $\pm$ 0.005 & 0.803 $\pm$ 0.004 & HPF$_1$ \\
    2458436.6146 & 36.2 $\pm$ 115.6 & 707.6 $\pm$ 803.3 & 90.6 $\pm$ 141.7 & 0.898 $\pm$ 0.004 & 0.838 $\pm$ 0.006 & 0.790 $\pm$ 0.005 & HPF$_1$ \\
    2458436.6186 & -116.9 $\pm$ 137.6 & 305.5 $\pm$ 814.0 & 84.5 $\pm$ 169.4 & 0.925 $\pm$ 0.004 & 0.887 $\pm$ 0.006 & 0.831 $\pm$ 0.005 & HPF$_1$ \\
    2458436.8202 & 322.8 $\pm$ 75.8 & 543.6 $\pm$ 512.2 & 14.5 $\pm$ 96.4 & 0.971 $\pm$ 0.004 & 0.935 $\pm$ 0.006 & 0.870 $\pm$ 0.004 & HPF$_1$ \\
    2458436.8241 & 347.9 $\pm$ 72.0 & 119.3 $\pm$ 688.4 & -76.4 $\pm$ 86.7 & 0.910 $\pm$ 0.004 & 0.871 $\pm$ 0.005 & 0.821 $\pm$ 0.004 & HPF$_1$ \\
    2458436.8280 & 452.5 $\pm$ 143.9 & 907.0 $\pm$ 866.1 & -133.9 $\pm$ 177.4 & 0.882 $\pm$ 0.005 & 0.855 $\pm$ 0.006 & 0.775 $\pm$ 0.005 & HPF$_1$ \\
    2458437.8224 & 308.3 $\pm$ 95.8 & 892.7 $\pm$ 301.8 & -111.5 $\pm$ 113.1 & 0.887 $\pm$ 0.005 & 0.813 $\pm$ 0.007 & 0.782 $\pm$ 0.006 & HPF$_1$ \\
    2458437.8263 & 332.5 $\pm$ 124.2 & 748.9 $\pm$ 714.4 & -183.9 $\pm$ 139.3 & 0.889 $\pm$ 0.004 & 0.847 $\pm$ 0.005 & 0.787 $\pm$ 0.004 & HPF$_1$ \\
    2458437.8303 & 237.5 $\pm$ 93.1 & 115.4 $\pm$ 721.8 & 40.3 $\pm$ 116.4 & 0.896 $\pm$ 0.004 & 0.849 $\pm$ 0.005 & 0.792 $\pm$ 0.004 & HPF$_1$ \\
    2458437.8343 & 331.4 $\pm$ 92.2 & 1069.8 $\pm$ 412.0 & -54.7 $\pm$ 114.5 & 1.048 $\pm$ 0.006 & 1.023 $\pm$ 0.008 & 0.925 $\pm$ 0.007 & HPF$_1$ \\
    2458451.7777 & -204.3 $\pm$ 113.8 & 990.0 $\pm$ 972.6 & -92.5 $\pm$ 139.6 & 0.910 $\pm$ 0.004 & 0.867 $\pm$ 0.005 & 0.815 $\pm$ 0.004 & HPF$_1$ \\
    2458451.7816 & -103.0 $\pm$ 155.9 & -590.1 $\pm$ 1007.4 & 358.8 $\pm$ 132.0 & 0.934 $\pm$ 0.004 & 0.887 $\pm$ 0.005 & 0.834 $\pm$ 0.004 & HPF$_1$ \\
    2458451.7856 & -428.4 $\pm$ 114.5 & 791.6 $\pm$ 1006.3 & 90.4 $\pm$ 139.7 & 0.888 $\pm$ 0.004 & 0.820 $\pm$ 0.005 & 0.785 $\pm$ 0.004 & HPF$_1$ \\
    2458454.7769 & 166.9 $\pm$ 101.7 & -474.6 $\pm$ 1299.3 & 191.3 $\pm$ 101.7 & 0.926 $\pm$ 0.005 & 0.864 $\pm$ 0.007 & 0.811 $\pm$ 0.006 & HPF$_1$ \\
    2458454.7808 & 92.0 $\pm$ 113.6 & 260.6 $\pm$ 933.4 & -11.8 $\pm$ 145.8 & 0.925 $\pm$ 0.005 & 0.892 $\pm$ 0.007 & 0.820 $\pm$ 0.006 & HPF$_1$ \\
    2458454.7848 & 235.3 $\pm$ 138.5 & 73.3 $\pm$ 569.6 & -30.7 $\pm$ 175.6 & 0.898 $\pm$ 0.004 & 0.843 $\pm$ 0.005 & 0.782 $\pm$ 0.004 & HPF$_1$ \\
    2458675.9378 & 16.3 $\pm$ 213.7 & -985.5 $\pm$ 1256.5 & 307.4 $\pm$ 245.5 & 1.055 $\pm$ 0.006 & 1.039 $\pm$ 0.009 & 0.951 $\pm$ 0.007 & HPF$_1$ \\
    2458675.9417 & 465.5 $\pm$ 170.0 & 292.4 $\pm$ 1314.9 & 48.0 $\pm$ 220.3 & 0.943 $\pm$ 0.004 & 0.871 $\pm$ 0.006 & 0.822 $\pm$ 0.005 & HPF$_1$ \\
    2458675.9456 & 101.6 $\pm$ 141.2 & 528.5 $\pm$ 351.9 & 250.5 $\pm$ 153.9 & 0.905 $\pm$ 0.004 & 0.859 $\pm$ 0.005 & 0.800 $\pm$ 0.004 & HPF$_1$ \\
    2458741.7631 & -288.4 $\pm$ 120.9 & 469.4 $\pm$ 790.5 & -88.9 $\pm$ 151.6 & 0.936 $\pm$ 0.004 & 0.895 $\pm$ 0.005 & 0.828 $\pm$ 0.004 & HPF$_1$ \\
    2458741.7669 & -193.6 $\pm$ 154.9 & -719.7 $\pm$ 1244.8 & -222.8 $\pm$ 177.5 & 0.934 $\pm$ 0.004 & 0.892 $\pm$ 0.006 & 0.834 $\pm$ 0.005 & HPF$_1$ \\
    2458741.7709 & -317.5 $\pm$ 147.8 & -320.3 $\pm$ 1059.5 & -91.0 $\pm$ 183.3 & 0.939 $\pm$ 0.005 & 0.893 $\pm$ 0.006 & 0.819 $\pm$ 0.005 & HPF$_1$ \\
    2458777.6734 & -592.2 $\pm$ 124.7 & 2669.3 $\pm$ 1018.0 & 89.2 $\pm$ 153.8 & 0.926 $\pm$ 0.004 & 0.868 $\pm$ 0.005 & 0.802 $\pm$ 0.004 & HPF$_1$ \\
    2458777.6773 & -548.6 $\pm$ 66.8 & 1594.7 $\pm$ 977.1 & 79.2 $\pm$ 78.2 & 0.900 $\pm$ 0.004 & 0.849 $\pm$ 0.005 & 0.796 $\pm$ 0.004 & HPF$_1$ \\
    2458777.6812 & -231.8 $\pm$ 104.7 & 1082.5 $\pm$ 1515.2 & -106.4 $\pm$ 126.6 & 0.867 $\pm$ 0.006 & 0.850 $\pm$ 0.008 & 0.765 $\pm$ 0.007 & HPF$_1$ \\
    2458872.6197 & -145.7 $\pm$ 155.1 & -746.9 $\pm$ 1406.9 & -309.3 $\pm$ 155.1 & 0.882 $\pm$ 0.006 & 0.849 $\pm$ 0.008 & 0.766 $\pm$ 0.006 & HPF$_1$ \\
    2458872.6237 & 292.3 $\pm$ 189.1 & 1433.5 $\pm$ 1734.1 & -375.9 $\pm$ 188.7 & 0.899 $\pm$ 0.004 & 0.855 $\pm$ 0.005 & 0.803 $\pm$ 0.004 & HPF$_1$ \\
    2458872.6276 & 397.6 $\pm$ 254.4 & -1401.2 $\pm$ 1603.1 & 246.2 $\pm$ 310.5 & 0.969 $\pm$ 0.005 & 0.948 $\pm$ 0.006 & 0.863 $\pm$ 0.005 & HPF$_1$ \\
    2459041.9448 & -83.1 $\pm$ 107.6 & -532.0 $\pm$ 899.8 & -40.6 $\pm$ 136.7 & 0.896 $\pm$ 0.004 & 0.872 $\pm$ 0.006 & 0.796 $\pm$ 0.004 & HPF$_1$ \\
    2459041.9488 & -69.4 $\pm$ 138.0 & 313.3 $\pm$ 706.2 & 64.5 $\pm$ 174.8 & 0.899 $\pm$ 0.006 & 0.835 $\pm$ 0.008 & 0.788 $\pm$ 0.007 & HPF$_1$ \\
    2459041.9527 & -220.0 $\pm$ 126.1 & 100.0 $\pm$ 1284.0 & -97.0 $\pm$ 158.9 & 0.903 $\pm$ 0.004 & 0.869 $\pm$ 0.005 & 0.802 $\pm$ 0.004 & HPF$_1$ \\
    2459046.9227 & 131.9 $\pm$ 86.2 & 146.7 $\pm$ 833.5 & 102.0 $\pm$ 100.9 & 0.909 $\pm$ 0.003 & 0.868 $\pm$ 0.004 & 0.800 $\pm$ 0.003 & HPF$_1$ \\
    2459046.9266 & 161.5 $\pm$ 142.5 & 1720.4 $\pm$ 934.7 & 347.5 $\pm$ 113.8 & 1.073 $\pm$ 0.006 & 1.023 $\pm$ 0.008 & 0.959 $\pm$ 0.006 & HPF$_1$ \\
    2459046.9305 & 141.9 $\pm$ 87.0 & -32.9 $\pm$ 852.3 & -7.3 $\pm$ 111.6 & 0.922 $\pm$ 0.005 & 0.907 $\pm$ 0.006 & 0.830 $\pm$ 0.005 & HPF$_1$ \\
    2459058.8913 & 264.0 $\pm$ 131.8 & 13.0 $\pm$ 746.9 & -54.5 $\pm$ 166.6 & 0.880 $\pm$ 0.005 & 0.828 $\pm$ 0.007 & 0.769 $\pm$ 0.006 & HPF$_1$ \\
    2459058.8953 & 99.3 $\pm$ 155.0 & 111.0 $\pm$ 605.8 & -126.8 $\pm$ 187.4 & 0.917 $\pm$ 0.004 & 0.864 $\pm$ 0.005 & 0.810 $\pm$ 0.004 & HPF$_1$ \\
    2459058.8993 & 61.5 $\pm$ 118.1 & -635.0 $\pm$ 465.8 & 202.7 $\pm$ 126.3 & 0.881 $\pm$ 0.003 & 0.839 $\pm$ 0.005 & 0.788 $\pm$ 0.004 & HPF$_1$ \\
    2459104.9842 & -330.7 $\pm$ 110.9 & 3389.0 $\pm$ 785.5 & 65.4 $\pm$ 142.2 & 0.906 $\pm$ 0.003 & 0.872 $\pm$ 0.004 & 0.804 $\pm$ 0.003 & HPF$_1$ \\
    2459104.9881 & -294.5 $\pm$ 137.2 & 1217.0 $\pm$ 519.8 & -79.1 $\pm$ 179.2 & 0.950 $\pm$ 0.005 & 0.908 $\pm$ 0.007 & 0.853 $\pm$ 0.006 & HPF$_1$ \\
    2459104.9921 & -177.1 $\pm$ 114.1 & 1614.4 $\pm$ 1058.4 & 214.8 $\pm$ 121.7 & 0.914 $\pm$ 0.006 & 0.859 $\pm$ 0.008 & 0.802 $\pm$ 0.007 & HPF$_1$ \\
    2459400.9444 & -50.7 $\pm$ 137.3 & 46.5 $\pm$ 1697.5 & -62.3 $\pm$ 175.9 & 0.885 $\pm$ 0.004 & 0.842 $\pm$ 0.005 & 0.787 $\pm$ 0.004 & HPF$_1$ \\
    2459400.9485 & -323.7 $\pm$ 78.7 & 511.6 $\pm$ 740.8 & -24.9 $\pm$ 101.4 & 0.990 $\pm$ 0.005 & 0.946 $\pm$ 0.006 & 0.874 $\pm$ 0.005 & HPF$_1$ \\
    2459400.9524 & -85.6 $\pm$ 114.8 & -324.9 $\pm$ 1476.3 & 6.0 $\pm$ 148.7 & 0.887 $\pm$ 0.004 & 0.836 $\pm$ 0.005 & 0.780 $\pm$ 0.004 & HPF$_1$ \\
    2459401.9486 & 82.8 $\pm$ 106.2 & -894.4 $\pm$ 1227.1 & 52.9 $\pm$ 134.6 & 0.885 $\pm$ 0.005 & 0.853 $\pm$ 0.006 & 0.782 $\pm$ 0.005 & HPF$_1$ \\
    2459401.9524 & 82.6 $\pm$ 105.9 & -57.7 $\pm$ 828.7 & -102.3 $\pm$ 128.9 & 0.894 $\pm$ 0.006 & 0.854 $\pm$ 0.009 & 0.785 $\pm$ 0.007 & HPF$_1$ \\
    2459401.9564 & 117.5 $\pm$ 62.4 & -671.6 $\pm$ 877.2 & -92.9 $\pm$ 69.8 & 0.913 $\pm$ 0.004 & 0.861 $\pm$ 0.005 & 0.800 $\pm$ 0.004 & HPF$_1$ \\
    2459411.9133 & -231.0 $\pm$ 136.8 & -895.4 $\pm$ 807.7 & -119.4 $\pm$ 173.4 & 0.914 $\pm$ 0.004 & 0.891 $\pm$ 0.006 & 0.820 $\pm$ 0.005 & HPF$_1$ \\
    2459411.9173 & -244.0 $\pm$ 80.8 & -185.5 $\pm$ 1436.3 & 98.3 $\pm$ 97.0 & 0.901 $\pm$ 0.004 & 0.867 $\pm$ 0.005 & 0.796 $\pm$ 0.004 & HPF$_1$ \\
    2459411.9213 & -237.8 $\pm$ 156.4 & 55.1 $\pm$ 1373.5 & 80.8 $\pm$ 201.3 & 0.912 $\pm$ 0.004 & 0.890 $\pm$ 0.005 & 0.813 $\pm$ 0.004 & HPF$_1$ \\
    2459423.8914 & -347.1 $\pm$ 244.4 & 4045.9 $\pm$ 1976.9 & 628.5 $\pm$ 191.1 & 0.896 $\pm$ 0.003 & 0.853 $\pm$ 0.004 & 0.784 $\pm$ 0.003 & HPF$_1$ \\
    2459423.8953 & -391.1 $\pm$ 137.3 & 2182.3 $\pm$ 1391.2 & -172.4 $\pm$ 166.6 & 0.926 $\pm$ 0.004 & 0.882 $\pm$ 0.006 & 0.821 $\pm$ 0.005 & HPF$_1$ \\
    2459423.8993 & -274.3 $\pm$ 113.3 & 1276.9 $\pm$ 1382.9 & 49.0 $\pm$ 147.7 & 0.917 $\pm$ 0.004 & 0.870 $\pm$ 0.005 & 0.811 $\pm$ 0.004 & HPF$_1$ \\
    2459426.8845 & -83.0 $\pm$ 96.8 & -1617.2 $\pm$ 1388.4 & 148.2 $\pm$ 109.4 & 0.885 $\pm$ 0.003 & 0.863 $\pm$ 0.005 & 0.785 $\pm$ 0.004 & HPF$_1$ \\
    2459426.8884 & -78.5 $\pm$ 128.4 & -558.8 $\pm$ 841.1 & -29.7 $\pm$ 162.7 & 0.898 $\pm$ 0.006 & 0.833 $\pm$ 0.008 & 0.792 $\pm$ 0.006 & HPF$_1$ \\
    2459426.8922 & 322.0 $\pm$ 107.1 & 764.7 $\pm$ 942.5 & 187.1 $\pm$ 116.6 & 0.896 $\pm$ 0.004 & 0.868 $\pm$ 0.005 & 0.801 $\pm$ 0.004 & HPF$_1$ \\
    \hline
    2459806.8532 & -97.8 $\pm$ 189.9 & -1311.9 $\pm$ 1378.7 & 234.4 $\pm$ 227.1 & 0.866 $\pm$ 0.006 & 0.839 $\pm$ 0.009 & 0.809 $\pm$ 0.007 & HPF$_2$ \\
    2459806.8572 & -87.3 $\pm$ 188.2 & -578.3 $\pm$ 933.8 & 57.0 $\pm$ 242.8 & 0.879 $\pm$ 0.005 & 0.839 $\pm$ 0.007 & 0.805 $\pm$ 0.006 & HPF$_2$ \\
    2459806.8612 & -6.1 $\pm$ 207.1 & 2178.7 $\pm$ 1154.6 & -17.9 $\pm$ 264.0 & 0.894 $\pm$ 0.005 & 0.853 $\pm$ 0.007 & 0.814 $\pm$ 0.006 & HPF$_2$ \\
    2459827.7790 & 166.7 $\pm$ 198.2 & 3433.1 $\pm$ 1527.6 & -15.7 $\pm$ 253.7 & 0.902 $\pm$ 0.004 & 0.862 $\pm$ 0.006 & 0.826 $\pm$ 0.005 & HPF$_2$ \\
    2459827.7831 & 187.1 $\pm$ 213.5 & 1850.9 $\pm$ 1244.2 & 150.3 $\pm$ 266.7 & 0.873 $\pm$ 0.004 & 0.831 $\pm$ 0.005 & 0.792 $\pm$ 0.004 & HPF$_2$ \\
    2459827.7870 & -29.1 $\pm$ 215.6 & 2862.2 $\pm$ 472.5 & 277.7 $\pm$ 250.5 & 0.872 $\pm$ 0.003 & 0.824 $\pm$ 0.004 & 0.798 $\pm$ 0.004 & HPF$_2$ \\
    2459828.7889 & 200.4 $\pm$ 134.9 & 734.7 $\pm$ 822.9 & -214.9 $\pm$ 148.3 & 0.975 $\pm$ 0.005 & 0.934 $\pm$ 0.006 & 0.897 $\pm$ 0.005 & HPF$_2$ \\
    2459828.7930 & 250.6 $\pm$ 96.1 & -945.7 $\pm$ 1387.6 & 80.4 $\pm$ 117.3 & 0.891 $\pm$ 0.004 & 0.852 $\pm$ 0.005 & 0.812 $\pm$ 0.004 & HPF$_2$ \\
    2459828.7969 & 473.0 $\pm$ 88.8 & -1934.0 $\pm$ 985.7 & 187.4 $\pm$ 84.1 & 0.980 $\pm$ 0.004 & 0.943 $\pm$ 0.006 & 0.893 $\pm$ 0.005 & HPF$_2$ \\
    2459843.9687 & 108.7 $\pm$ 145.1 & 153.8 $\pm$ 1164.0 & 137.5 $\pm$ 179.4 & 0.979 $\pm$ 0.005 & 0.953 $\pm$ 0.006 & 0.901 $\pm$ 0.005 & HPF$_2$ \\
    2459843.9726 & 503.7 $\pm$ 104.7 & -648.3 $\pm$ 930.0 & 15.3 $\pm$ 133.8 & 0.878 $\pm$ 0.004 & 0.845 $\pm$ 0.006 & 0.799 $\pm$ 0.005 & HPF$_2$ \\
    2459843.9766 & 443.3 $\pm$ 120.7 & -66.9 $\pm$ 1021.6 & -177.8 $\pm$ 137.5 & 0.901 $\pm$ 0.004 & 0.860 $\pm$ 0.006 & 0.818 $\pm$ 0.005 & HPF$_2$ \\
    2459886.6248 & -338.5 $\pm$ 173.9 & 986.6 $\pm$ 1477.8 & 4.2 $\pm$ 222.7 & 0.899 $\pm$ 0.005 & 0.870 $\pm$ 0.007 & 0.811 $\pm$ 0.006 & HPF$_2$ \\
    2459886.6287 & -367.5 $\pm$ 147.8 & 85.6 $\pm$ 1266.4 & 125.4 $\pm$ 179.0 & 0.888 $\pm$ 0.003 & 0.846 $\pm$ 0.005 & 0.810 $\pm$ 0.004 & HPF$_2$ \\
    2459886.6327 & -385.8 $\pm$ 92.5 & -794.3 $\pm$ 967.1 & -10.9 $\pm$ 115.9 & 0.886 $\pm$ 0.004 & 0.833 $\pm$ 0.005 & 0.797 $\pm$ 0.004 & HPF$_2$ \\
    2459890.6280 & -252.9 $\pm$ 84.2 & -397.4 $\pm$ 780.5 & -130.9 $\pm$ 91.9 & 0.881 $\pm$ 0.004 & 0.852 $\pm$ 0.006 & 0.802 $\pm$ 0.005 & HPF$_2$ \\
    2459890.6320 & -250.1 $\pm$ 84.7 & -431.2 $\pm$ 843.3 & -107.1 $\pm$ 97.4 & 0.859 $\pm$ 0.003 & 0.813 $\pm$ 0.005 & 0.784 $\pm$ 0.004 & HPF$_2$ \\
    2459890.6360 & -252.1 $\pm$ 122.3 & -1085.1 $\pm$ 1131.8 & -139.2 $\pm$ 142.8 & 0.893 $\pm$ 0.005 & 0.863 $\pm$ 0.007 & 0.836 $\pm$ 0.006 & HPF$_2$ \\
    2459913.5496 & 64.0 $\pm$ 119.9 & -2167.1 $\pm$ 1539.5 & -47.9 $\pm$ 153.7 & 0.902 $\pm$ 0.004 & 0.859 $\pm$ 0.005 & 0.813 $\pm$ 0.004 & HPF$_2$ \\
    2459913.5538 & 52.4 $\pm$ 245.7 & -432.6 $\pm$ 1550.5 & 85.9 $\pm$ 312.8 & 0.894 $\pm$ 0.004 & 0.840 $\pm$ 0.005 & 0.811 $\pm$ 0.004 & HPF$_2$ \\
    2459913.5577 & 34.5 $\pm$ 178.7 & -1463.2 $\pm$ 2104.6 & -283.8 $\pm$ 198.2 & 0.902 $\pm$ 0.004 & 0.850 $\pm$ 0.006 & 0.825 $\pm$ 0.005 & HPF$_2$ \\
    \enddata
    \end{deluxetable*}

\section{Results of Models 1 and 2 Fit\label{sec:phase_curve_and_posteriors}}
\begin{figure*}[!ht]
        \centering
        \includegraphics[width=0.89\linewidth]{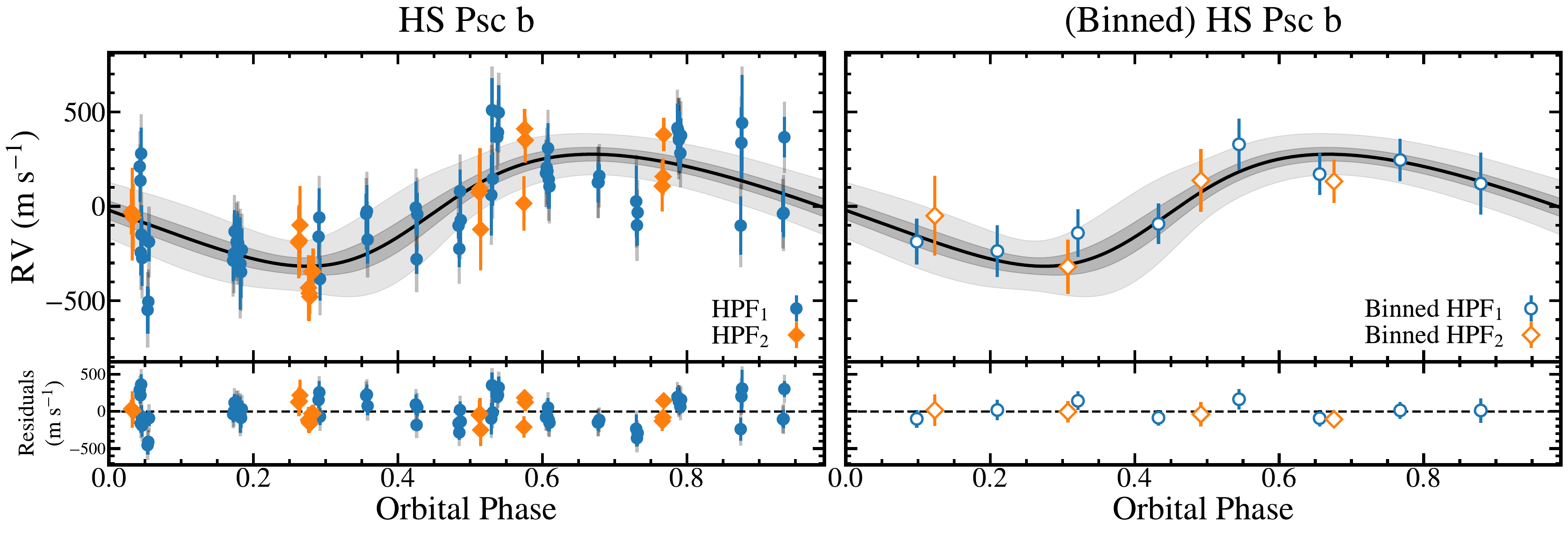}
        
        \hspace{-3mm}
        \includegraphics[width=0.925\linewidth]{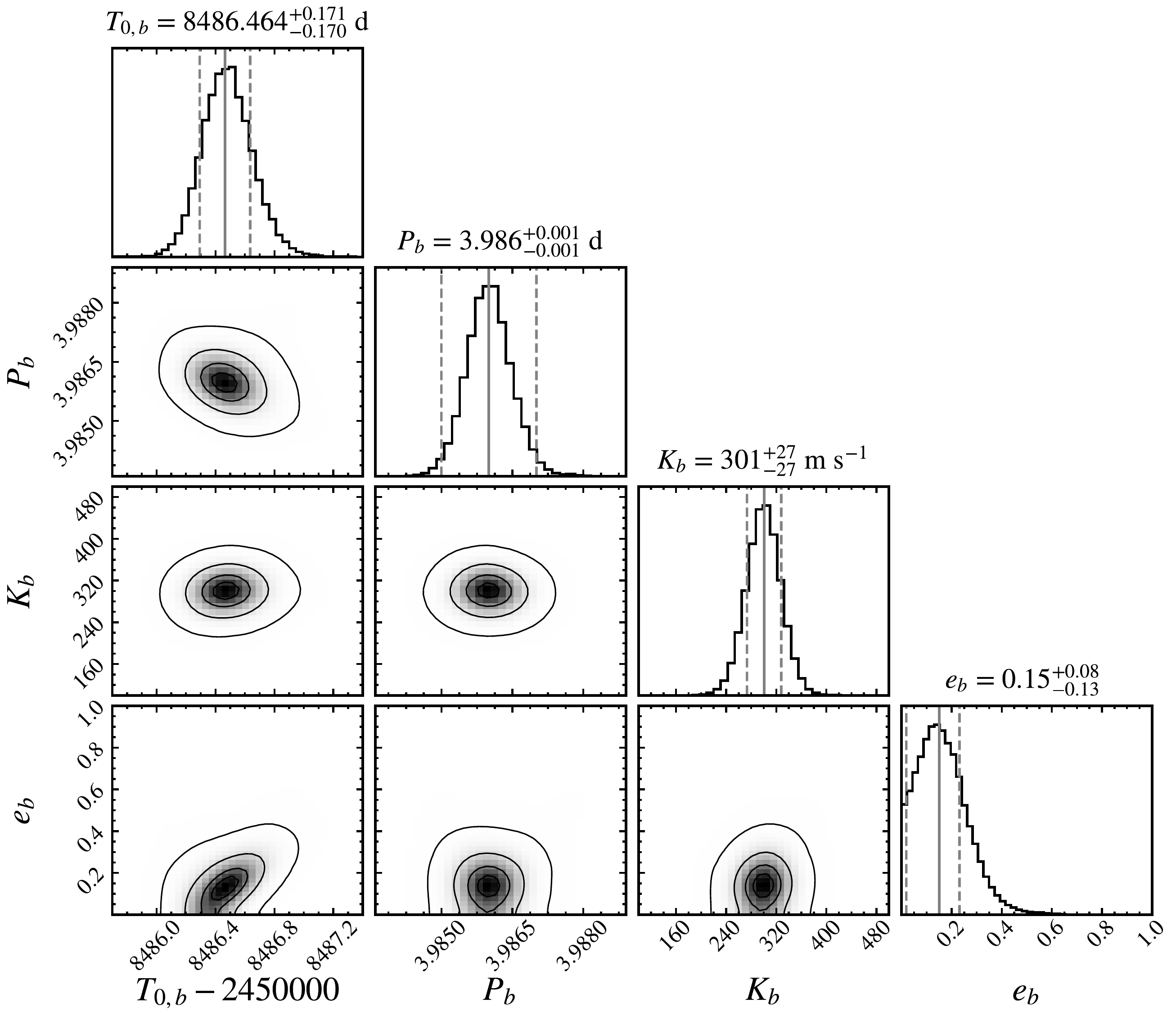}
	    \caption{The same plots as \autoref{fig:model_3_results} for the Model 1 (Keplerian-only) fits.}
	\label{fig:model_1_results}
\end{figure*}

\begin{figure*}[!ht]
        \centering
        \includegraphics[width=0.89\linewidth]{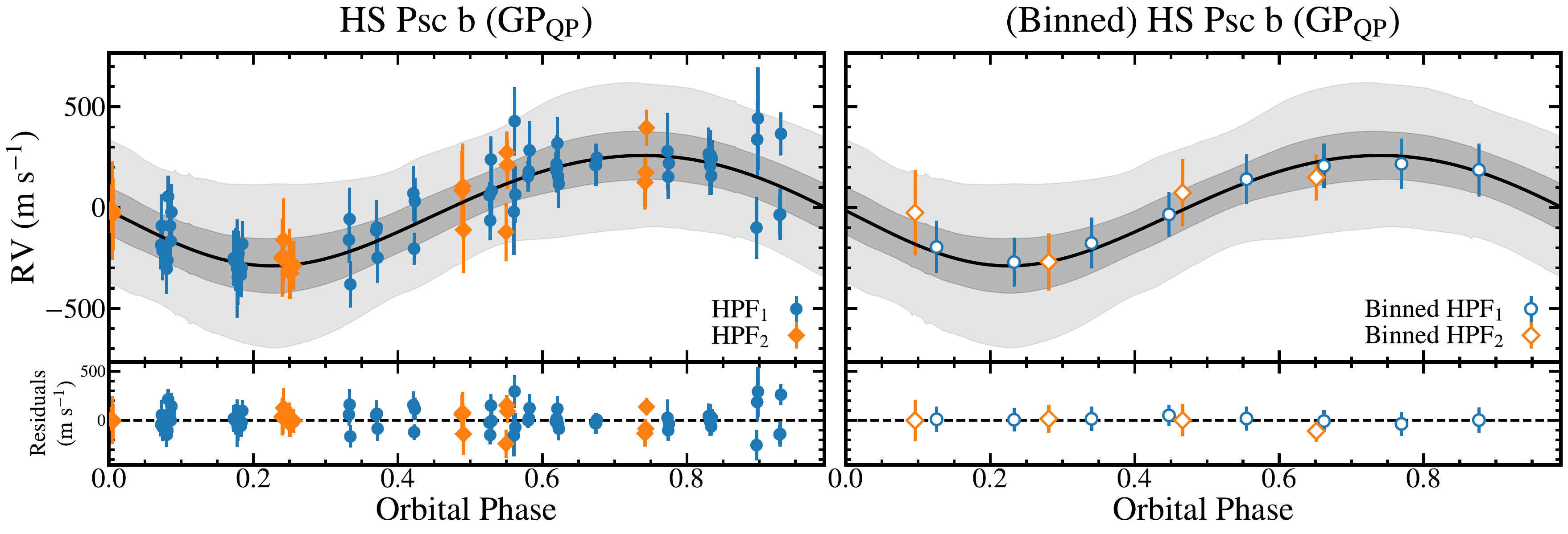}
        
        \hspace{-3mm}
        \includegraphics[width=0.925\linewidth]{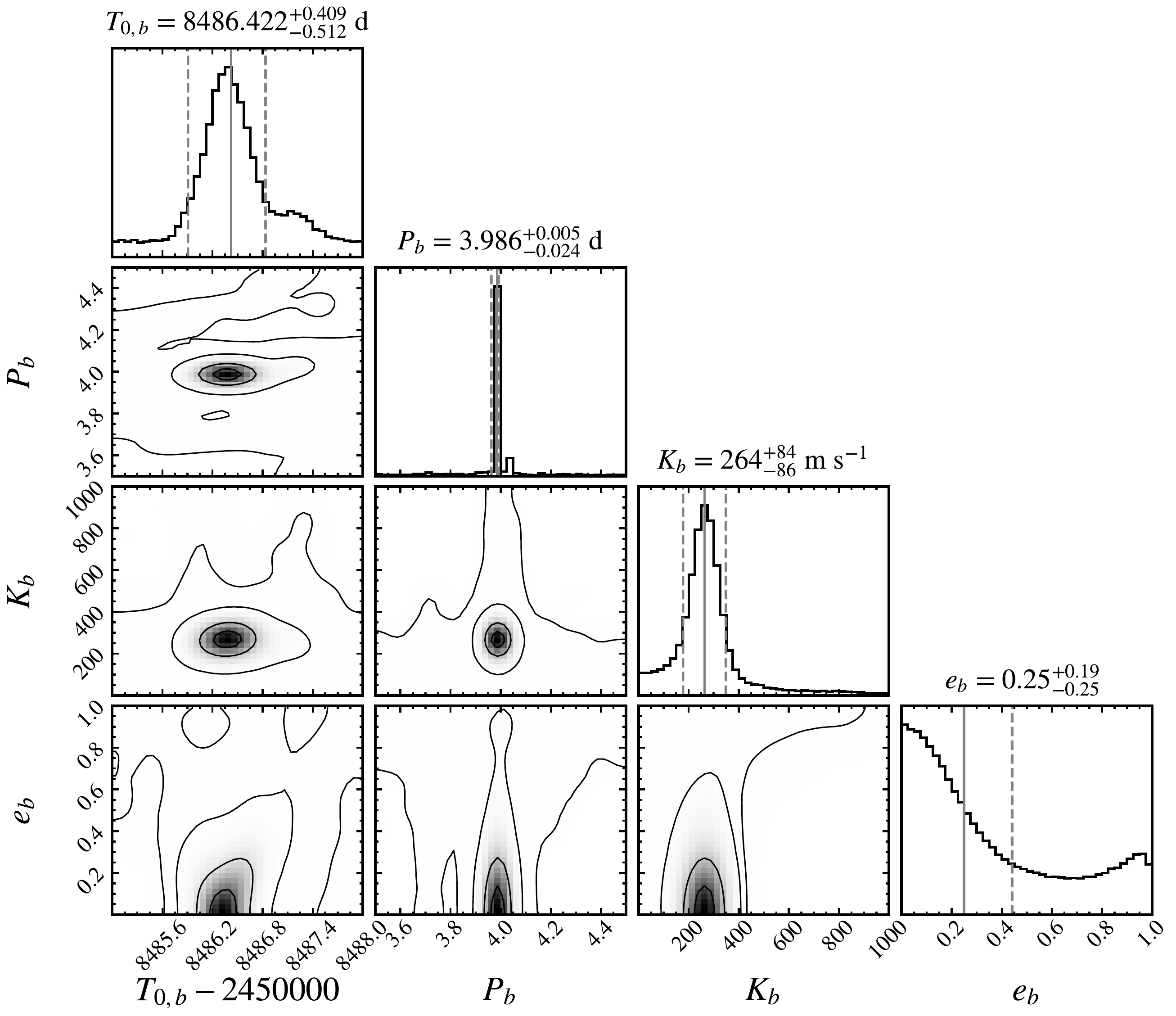}
	    \caption{The same plots as \autoref{fig:model_3_results} for the Model 2 (Keplerian and QP GP) fits.}
	\label{fig:model_2_results}
\end{figure*}

\Cref{tab:model_params,tab:model_params_gp_qp} summarize the prior choices and results of the Models 1 and 2 fit, respectively. \Cref{fig:model_1_results,fig:model_2_results} display the best-fit phased RV curve and posterior distributions of the planetary parameters Models 1 and 2, respectively. See \Cref{sec:model_1,sec:model_2} for details.

\begin{deluxetable*}{lccc}[b]
    \setlength{\tabcolsep}{24pt}
    \tablecaption{\label{tab:model_params}Parameter Priors and Posteriors from Keplerian-only Fit (Model 1) to HPF RVs of HS Psc.}
    \tablehead{\colhead{Parameter} & \colhead{Prior} & \colhead{MAP} & \colhead{Median $\pm$ 1$\sigma$}}
    \startdata
    \multicolumn{4}{c}{Fitted Parameters} \\
    \hline
    \multicolumn{4}{c}{Keplerian and Instrumental Parameters} \\
    $P_b$ (d) & $\mathcal{U}[0.5, 10.0]$ & $3.986$ & $3.986_{-0.001}^{+0.001}$ \\
    $T_{0, b}$ (d) & $\mathcal{U}[2458483.0, 2458487.5]$ & $2458486.494$ & $2458486.464_{-0.170}^{+0.171}$ \\
    $K_b$ (m s$^{-1}$) & $\mathcal{U}[1.0, 1000.0]$ & $296.5$ & $301 \pm 27$ \\
    $\sqrt{e_b} \sin{\omega_*}$ & $\mathcal{U}[-1, 1]$ & $-0.39$ & $-0.20_{-0.24}^{+0.22}$ \\
    $\sqrt{e_b} \cos{\omega_*}$ & $\mathcal{U}[-1, 1]$ & $-0.17$ & $-0.22_{-0.22}^{+0.18}$ \\
    $\sigma_{\mathrm{HPF}_{1}}$ (m s$^{-1}$) & $\mathcal{J}(1, 100)$ & $155.9$ & $161.3_{-24.6}^{+21.0}$ \\
    $\sigma_{\mathrm{HPF}_{2}}$ (m s$^{-1}$) & $\mathcal{J}(1, 100)$ & $0.0$ & $17.4_{-17.4}^{+27.2}$ \\
    $\gamma_{\mathrm{HPF}_{1}}$ (km s$^{-1}$) & $\mathcal{U}[-1.0922, 0.9655]$ & $-0.043$ & $-0.037_{-0.026}^{+0.025}$ \\
    $\gamma_{\mathrm{HPF}_{2}}$ (km s$^{-1}$) & $\mathcal{U}[-0.8858 , 1.0037]$ & $0.093$ & $0.083_{-0.035}^{+0.037}$ \\
    \hline
    \multicolumn{4}{c}{Derived Parameters} \\
    \hline
    $m_b \sin i$ $(M_\mathrm{Jup})$ & $\cdots$ & $1.76$ & $1.78_{-0.21}^{+0.22}$ \\
    $a_b$ (AU) & $\cdots$ & $0.043$ & $0.043_{-0.002}^{+0.002}$ \\
    $T_\mathrm{peri, b}$ (d) & $\cdots$ & $2458488.13$ & $2458487.56_{-0.41}^{+0.79}$ \\
    $e_b$ & $\cdots$ & $0.18$ & $0.15_{-0.13}^{+0.08}$ \\
    $\omega_*$ $(\degree)$ & $\cdots$ & $247.19$ & $218.98_{-37.05}^{+58.57}$ \\
    \enddata
\end{deluxetable*}

\begin{deluxetable*}{lccc}
    \setlength{\tabcolsep}{16pt}
    \tablecaption{\label{tab:model_params_gp_qp}Parameter Priors and Posteriors from Keplerian and Quasi-Periodic GP Fit (Model 2) to HPF RVs of HS Psc.}
    \tablehead{\colhead{Parameter} & 
    \colhead{Prior} & \colhead{MAP} & \colhead{Median $\pm$ 1$\sigma$}}
    \startdata
    \multicolumn{4}{c}{{Fitted Parameters}} \\
    \hline
    \multicolumn{4}{c}{Keplerian and Instrumental Parameters} \\
    $P_b$ (d) & $\mathcal{U}[3.5, 4.5]$ & $3.987$ & $3.986_{-0.024}^{+0.005}$ \\
    $T_{0, b}$ (d) & $\mathcal{U}[2458485.0, 2458488.0]$ & $2458486.333$ & $2458486.422_{-0.512}^{+0.409}$ \\
    $K_b$ (m s$^{-1}$) & $\mathcal{U}[1.0, 1000.0]$ & $273.9$ & $264_{-86}^{+84}$ \\
    $\sqrt{e_b} \sin{\omega_*}$ & $\mathcal{U}[-1, 1]$ & $0.09$ & $0.04_{-0.48}^{+0.43}$ \\
    $\sqrt{e_b} \cos{\omega_*}$ & $\mathcal{U}[-1, 1]$ & $-0.23$ & $-0.11_{-0.37}^{+0.38}$ \\
    $\sigma_{\mathrm{HPF}_{1}}$ (m s$^{-1}$) & $\mathcal{J}(1, 100)$ & $0.1$ & $8.0_{-8.0}^{+10.7}$ \\
    $\sigma_{\mathrm{HPF}_{2}}$ (m s$^{-1}$) & $\mathcal{J}(1, 100)$ & $0.1$ & $8.1_{-8.1}^{+11.7}$ \\
    $\gamma_{\mathrm{HPF}_{1}}$ (km s$^{-1}$) & $\mathcal{U}[-1.0922, 0.9655]$ & $-0.031$ & $-0.040_{-0.052}^{+0.050}$ \\
    $\gamma_{\mathrm{HPF}_{2}}$ (km s$^{-1}$) & $\mathcal{U}[-0.8858 , 1.0037]$ & $0.050$ & $0.065_{-0.089}^{+0.094}$ \\
    \multicolumn{4}{c}{Quasi-Periodic Kernel Hyperparameters} \\
    $A$ (m s$^{-1}$) & $\mathcal{U}[0.0, 1000.0]$ & $141.2$ & $185.2_{-54.6}^{+40.1}$ \\
    $\lambda_e$ & $\mathcal{U}[0.01, 5.0]$ & $3.82$ & $2.07_{-2.06}^{+1.07}$ \\
    $\lambda_p$ & $\mathcal{U}[0.01, 5.0]$ & $0.25$ & $0.30_{-0.27}^{+0.14}$ \\
    $P_\mathrm{GP}$ & $\mathcal{N}[1.09, 0.02]$ & $1.08$ & $1.09_{-0.02}^{+0.02}$ \\
    \hline
    \multicolumn{4}{c}{Derived Parameters} \\
    \hline
    $m_b \sin i$ $(M_\mathrm{Jup})$ & $\cdots$ & $1.78$ & $1.46_{-0.43}^{+0.54}$ \\
    $a_b$ (AU) & $\cdots$ & $0.044$ & $0.044_{-0.002}^{+0.002}$ \\
    $T_\mathrm{peri, b}$ (d) & $\cdots$ & $2458487.03$ & $2458486.78_{-0.93}^{+1.35}$ \\
    $e_b$ & $\cdots$ & $0.06$ & $0.25_{-0.25}^{+0.19}$ \\
    $\omega_*$ $(\degree)$ & $\cdots$ & $159.55$ & $170.36_{-100.95}^{+99.69}$ \\
    \enddata
\end{deluxetable*}

\bibliography{hs_psc}{}
\bibliographystyle{aasjournal}

\end{document}